\documentclass{emulateapj}
\usepackage{apjfonts}
\usepackage{multirow}

\shorttitle{HST UV Observations of SNe Ia} \shortauthors{Wang et al.}

\begin{document}
\title{Evidence for Type Ia Supernova Diversity \\ from Ultraviolet Observations
with the Hubble Space Telescope}

\author{Xiaofeng Wang\altaffilmark{1,2,3}, Lifan Wang\altaffilmark{2}, Alexei V. Filippenko\altaffilmark{3},
Eddie Baron\altaffilmark{4}, Markus Kromer\altaffilmark{5}, Dennis Jack\altaffilmark{6}, Tianmeng Zhang\altaffilmark{7},
Greg Aldering\altaffilmark{8}, Pierre Antilogus\altaffilmark{9}, W. David Arnett\altaffilmark{10}, Dietrich Baade\altaffilmark{11},
Brian J. Barris\altaffilmark{12}, Stefano Benetti\altaffilmark{13}, Patrice Bouchet\altaffilmark{14},
Adam S. Burrows\altaffilmark{15}, Ramon Canal\altaffilmark{16}, Enrico Cappellaro\altaffilmark{13}, Raymond Carlberg\altaffilmark{17},
Elisa di Carlo\altaffilmark{18}, Peter Challis\altaffilmark{19}, Arlin Crotts\altaffilmark{20}, John I. Danziger\altaffilmark{21},
Massimo Della Valle\altaffilmark{22,23},  Michael Fink\altaffilmark{24}, Ryan J. Foley\altaffilmark{19, 25},
Claes Fransson\altaffilmark{26}, Avishay Gal-Yam\altaffilmark{27}, Peter Garnavich\altaffilmark{28}, Chris L. Gerardy\altaffilmark{29},
Gerson Goldhaber\altaffilmark{8}, Mario Hamuy\altaffilmark{30}, Wolfgang Hillebrandt\altaffilmark{5}, Peter H\"{o}flich\altaffilmark{29},
Stephen T. Holland\altaffilmark{31}, Daniel E. Holz\altaffilmark{32}, John P. Hughes\altaffilmark{33},
David J. Jeffery\altaffilmark{34}, Saurabh W. Jha\altaffilmark{33}, Dan Kasen\altaffilmark{35},
Alexei M. Khokhlov\altaffilmark{32}, Robert P. Kirshner\altaffilmark{19}, Robert Knop\altaffilmark{36},
Cecilia Kozma\altaffilmark{26}, Kevin Krisciunas\altaffilmark{2},
Brian C. Lee\altaffilmark{8}, Bruno Leibundgut\altaffilmark{37}, Eric J. Lentz\altaffilmark{38},
Douglas C. Leonard\altaffilmark{39}, Walter H. G. Lewin\altaffilmark{40}, Weidong Li\altaffilmark{3},
Mario Livio\altaffilmark{41}, Peter Lundqvist\altaffilmark{26}, Dan Maoz\altaffilmark{42},
Thomas Matheson\altaffilmark{43}, Paolo Mazzali\altaffilmark{13,5}, Peter Meikle\altaffilmark{44},
Gajus Miknaitis\altaffilmark{45}, Peter Milne\altaffilmark{10}, Stefan Mochnacki\altaffilmark{46},
Ken'ichi Nomoto\altaffilmark{47}, Peter E. Nugent\altaffilmark{8}, Elaine Oran\altaffilmark{47},
Nino Panagia\altaffilmark{41}, Saul Perlmutter\altaffilmark{8},
Mark M. Phillips\altaffilmark{49}, Philip Pinto\altaffilmark{10}, Dovi Poznanski\altaffilmark{50},
Christopher J. Pritchet\altaffilmark{51}, Martin Reinecke\altaffilmark{5}, Adam G. Riess\altaffilmark{41},
Pilar Ruiz-Lapuente\altaffilmark{16}, Richard Scalzo\altaffilmark{8}, Eric M. Schlegel\altaffilmark{52},
Brian P. Schmidt\altaffilmark{53}, James Siegrist\altaffilmark{8}, Alicia M. Soderberg\altaffilmark{19},
Jesper Sollerman\altaffilmark{26}, George Sonneborn\altaffilmark{30}, Anthony Spadafora\altaffilmark{8},
Jason Spyromilio\altaffilmark{37}, Richard A. Sramek\altaffilmark{54}, Sumner G. Starrfield\altaffilmark{55},
Louis G. Strolger\altaffilmark{56}, Nicholas B. Suntzeff\altaffilmark{2}, Rollin Thomas\altaffilmark{8},
John L. Tonry\altaffilmark{12}, Amedeo Tornambe\altaffilmark{57}, James W. Truran\altaffilmark{32},
Massimo Turatto\altaffilmark{21}, Michael Turner\altaffilmark{32}, Schuyler D. Van Dyk\altaffilmark{58},
Kurt Weiler\altaffilmark{48}, J. Craig Wheeler\altaffilmark{59}, Michael Wood-Vasey\altaffilmark{60},
Stanford E. Woosley\altaffilmark{61}, and Hitoshi Yamaoka\altaffilmark{62}}

\altaffiltext{1}{Physics Department and Tsinghua Center for Astrophysics (THCA), Tsinghua University, Beijing, 100084, China; wang\_xf@mail.tsinghua.edu.cn.}
\altaffiltext{2}{Physics and Astronomy Department, Texas A\&M University, College Station, TX 77843, USA.}
\altaffiltext{3}{Department of Astronomy, University of California, Berkeley, CA 94720-3411, USA.}
\altaffiltext{4}{Department of Physics, University of Oklahoma, Norman, OK 73019, USA.}
\altaffiltext{5}{Max-Planck-Institut f\"{u}r Astrophysik, Karl-Schwarzschild-Str. 1, 85748 Garching, Germany.}
\altaffiltext{6}{Hamburger Sternwarte, Gojenbergsweg 112, 21029 Hamburg, Germany.}
\altaffiltext{7}{National Astronomical Observatory of China, Chinese Academy of Sciences, Beijing, 100012, China.}
\altaffiltext{8}{Lawrence Berkeley National Laboratory, Berkeley, CA 94720, USA.}
\altaffiltext{9}{Laboratoire de Physique Nucleaire des Hautes Energies, Paris, France.}
\altaffiltext{10}{Steward Observatory, University of Arizona, Tucson, AZ 85721, USA.}
\altaffiltext{11}{European Southern Observatory, 85748, Garching bei M\"{u}nchen, Germany.}
\altaffiltext{12}{Institute for Astronomy, University of Hawaii, Honolulu, HI 96822, USA.}
\altaffiltext{13}{Osservatorio Astronomico di Padova, 35122 Padova, Italy.}
\altaffiltext{14}{CEA/DSM/DAPNIA/Service d'Astrophysique, 91191 Gif-sur-Yvette Cedex, France.}
\altaffiltext{15}{Department of Astrophysical Sciences, Princeton University, Princeton, NJ 08544, USA.}
\altaffiltext{16}{Universidad de Barcelona, Barcelona 8007, Spain.}
\altaffiltext{17}{University of Toronto, Toronto, ON M5S 3J3, Canada.}
\altaffiltext{18}{INAF, Osservatorio Astronomico di Teramo, 64100 Teramo, Italy.}
\altaffiltext{19}{Harvard/Smithsonian Center Astrophysics, Cambridge, MA 02138, USA.}
\altaffiltext{20}{Department of Astronomy, Columbia University, New York, NY 10025, USA.}
\altaffiltext{21}{INAF, Osservatorio Astronomico di Trieste, I-34143, Trieste, Italy.}
\altaffiltext{22}{Capodimonte Astronomical Observatory, INAF-Napoli, I-80131, Napoli, Italy.}
\altaffiltext{23}{International Center for Relativistic Astrophysics, I-65122, Pescara, Italy.}
\altaffiltext{24}{Institut f\"{u}r Theoretische Physik und Astrophysik,
Universit\"{a}t W\"{u}rzburg, Am Hubland, D-97074 W\"{u}rzburg, Germany.}
\altaffiltext{25}{Clay Fellow.}
\altaffiltext{26}{Stockholm University, SE-106 91 Stockholm, Sweden.}
\altaffiltext{27}{Weizmann Institute of Science, Rehovot, 76100, Israel.}
\altaffiltext{28}{University of Notre Dame, Notre Dame, IN 46556, USA.}
\altaffiltext{29}{Florida State University, Tallahassee, FL 32306, USA.}
\altaffiltext{30}{Universidad de Chile, Casilla 36D, Santiago, Chile.}
\altaffiltext{31}{Laboratory for Observational Cosmology, NASA Goddard Space Flight Center, Code 665, Greenbelt, MD 20771, USA.}
\altaffiltext{32}{University of Chicago, Chicago, IL 60637, USA.}
\altaffiltext{33}{Department of Physics and Astronomy, Rutgers, the State University of New Jersey, Piscataway, NJ 08854, USA.}
\altaffiltext{34}{Northern Arizona University, Flagstaff, AZ 86011, USA.}
\altaffiltext{35}{Department of Physics, University of California, Berkeley, CA 94720, USA.}
\altaffiltext{36}{Quest University Canada, Squamish, BC, Canada.}
\altaffiltext{37}{European Southern Observatory, 85748 Garching bei M\"{u}nchen,  Germany.}
\altaffiltext{38}{Department of Physics and Astronomy, University of Tennessee, Knoxville, TN 37996, USA.}
\altaffiltext{39}{Department of Astronomy, San Diego State University, San Diego, CA 92182, USA.}
\altaffiltext{40}{Massachusetts Institute of Technology, Cambridge, MA 02139, USA.}
\altaffiltext{41}{Space Telescope Science Institute, Baltimore, MD 21218, USA.}
\altaffiltext{42}{Tel Aviv University, Wise Observatory, 69978, Tel Aviv, Israel.}
\altaffiltext{43}{National Optical Astronomy Observatory, Tucson, AZ 85719, USA.}
\altaffiltext{44}{Imperial College of Science Technology and Medicine, London SW7 2AZ, UK.}
\altaffiltext{45}{Center for Neighborhood Technology, 2125 W. North Ave., Chicago IL 60647}
\altaffiltext{46}{University of Toronto, Toronto, ON M5S 3G4, Canada.}
\altaffiltext{47}{University of Tokyo, IPMU, Kashiwa, Chiba 277-8583, Japan.}
\altaffiltext{48}{Naval Research Laboratory, Washington, DC 20375, USA.}
\altaffiltext{49}{Carnegie Institution of Washington, Washington, DC 20005, USA.}
\altaffiltext{50}{School of Physics and Astronomy, Tel-Aviv University, Tel Aviv 69978, Israel.}
\altaffiltext{51}{University of Victoria, Victoria, BC V8W 2Y2, Canada.}
\altaffiltext{52}{Physics \& Astronomy Department, University of Texas at San Antonio, San Antonio, TX 78249, USA.}
\altaffiltext{53}{Australian National University, Canberra ACT 0200, Australia.}
\altaffiltext{54}{Associated Universities, Inc., Washington, DC 20036, USA.}
\altaffiltext{55}{Arizona State University, Tempe, AZ 85287, USA.}
\altaffiltext{56}{Western Kentucky University, Bowling Green, KY 42101, USA.}
\altaffiltext{57}{INAF, Rome Astronomical Observatory, 00136 Roma, Italy.}
\altaffiltext{58}{IPAC, California Institute of Technology, Pasadena, CA 91125, USA.}
\altaffiltext{59}{Department of Astronomy and McDonald Observatory, University of Texas, Austin, TX 78712, USA.}
\altaffiltext{60}{Pittsburgh Particle Physics, Astrophysics, and Cosmology Center (Pitt-PACC),
University of Pittsburgh, Pittsburgh, PA 15260, USA.}
\altaffiltext{61}{University of California, Santa Cruz, CA 95060, USA.}
\altaffiltext{62}{Graduate School of Sciences, Kyushu University, Fukuoka 812-8581, Japan.}

\begin{abstract}
  We present ultraviolet (UV) spectroscopy and photometry of four Type
  Ia supernovae (SNe 2004dt, 2004ef, 2005M, and 2005cf) obtained with
  the UV prism of the Advanced Camera for Surveys on the {\it Hubble
    Space Telescope}. This dataset provides unique spectral time
  series down to 2000~\AA. Significant diversity is seen in the
  near-maximum-light spectra ($\sim 2000$--3500~\AA) for this small
  sample. The corresponding photometric data, together with archival
  data from {\it Swift} Ultraviolet/Optical Telescope observations,
  provide further evidence of increased dispersion in the UV emission
  with respect to the optical. The peak luminosities measured in the
  uvw1/F250W filter are found to correlate with the $B$-band light-curve
  shape parameter $\Delta m_{15}(B)$, but with much larger scatter
  relative to the correlation in the broad-band $B$ band (e.g., $\sim
  0.4$ mag versus $\sim 0.2$ mag for those with $0.8 < \Delta m_{15}(B) <
  1.7$ mag). SN 2004dt is found as an outlier of this correlation (at
  $> 3\sigma$), being brighter than normal SNe~Ia such as SN 2005cf by
  $\sim 0.9$ mag and $\sim 2.0$ mag in the uvw1/F250W and uvm2/F220W filters,
  respectively. We show that different progenitor metallicity or
  line-expansion velocities alone cannot explain such a large
  discrepancy. Viewing-angle effects, such as due to an asymmetric
  explosion, may have a significant influence on the flux emitted in
  the UV region. Detailed modeling is needed to disentangle and
  quantify the above effects.
\end{abstract}


\section{Introduction}

The utility of Type Ia supernovae (SNe~Ia; see Filippenko 1997 for a
review of SN spectral classification) as cosmological probes
depends on the degree of our understanding of SN~Ia physics, and on
various systematic effects such as cosmic chemical evolution.  There
is increasing evidence showing that even the so-called
``Branch-normal'' SNe~Ia (Branch et al. 1993) exhibit diversity in
their spectral features and light-curve shapes that do not correlate
with light-curve parameters such as the decline rate
(Phillips 1993; Benetti et al. 2005; Branch et al. 2009; Wang et
al. 2009a; Zhang et al. 2010; H\"{o}flich et al. 2010).  For example,
it has been recently shown that SNe~Ia having the same light-curve
shape parameter such as $\Delta m_{15}(B)$ (Phillips 1993) but faster
expanding ejecta are on average $\sim 0.1$ mag redder in $B-V$
color near maximum light (Wang et al. 2009a, hereafter W09a).
Possible origins of such a color difference include a change of the
dust obscuring the SN (e.g., circumstellar dust versus
interstellar dust; W09a), an effect of line blanketing (Foley \& Kasen
2011), or a projection effect in an asymmetric explosion (Maeda et
al. 2011).  Moreover, the peak luminosity of SNe~Ia seems to show an
additional dependence on the global characteristics of their host
galaxies: events of the same light-curve shape and color may be
$\sim 0.1$ mag brighter in massive (presumably metal-rich) host galaxies
(Kelly et al. 2010; Sullivan et al. 2010; Lampeitl et al. 2010).
This correlation might also indicate
variations in the explosion physics and/or differences of the
progenitors in a range of environments.

Ultraviolet (UV) observations are of particular importance in
understanding both the diversity of SNe~Ia and their physical
evolution from low to high redshifts. In conjunction with existing
optical and near-infrared data, UV observations improve the
determination of the bolometric light curves of SNe~Ia. The UV
spectrum at early times forms mainly in the outer shells of the
explosion ejecta where the unburned outer layers of the white dwarf
play a large role in shaping the appearance of the spectrum. The
dependence of the UV emission on the progenitor metallicity has been
theoretically studied by several authors (Pauldrach et al. 1996;
H\"{o}flich et al. 1998; Mazzali et al. 2000; Lentz et al. 2000;
Timmes et al. 2003; Sauer et al. 2008).  Both H\"{o}flich et
al. (1998) and Sauer et al. (2008) argued that increasing the
metallicity in the outer layers would lead to a stronger UV flux, with
the latter highlighting the contribution from the reverse-fluorescence
scattering of photons from red to blue wavelengths. In contrast to the
above studies, Lentz et al. (2000) predicted that increasing the
metallicity of the progenitor should cause a decrease in the level of
the UV pseudocontinuum due to a decrease in the line opacity and
cooling. A much larger effect was proposed by Timmes et al. (2003),
who argued for a significant increase of $^{56}$Ni, and hence of the
resulting luminosity, with increasing metallicity (but see Howell et
al. 2009). We note that current theories do not offer a consensus
picture as to the effects of changing metallicity on the overall UV
flux.  UV observations of a large sample of SNe~Ia are needed to help
quantify the UV properties and constrain the above models.

The necessity of observing above the Earth's atmosphere has resulted
in a sparse UV dataset for low-redshift SNe~Ia; see Panagia (2003,
2007) and Foley, Filippenko, \& Jha (2008a) for summaries of {\it
  International Ultraviolet Explorer (IUE)} and {\it Hubble Space
  Telescope (HST)} spectra. Recently, the {\it Swift} satellite
obtained UV photometry of nearby SNe~Ia (Milne et al. 2010; Brown et
al. 2010), as well as some UV grism spectra (Bufano et~al. 2009; Foley
et al.  2011). With the successful repair of the Space Telescope
Imaging Spectrograph (STIS) onboard {\it HST} during the 2009
Servicing Mission 4, obtaining UV spectra with STIS has also become
possible (Cooke et al. 2011). Observations of distant SNe~Ia (with
ground-based telescopes) provide larger samples that probe the UV
spectral properties at high redshift (e.g., Ellis et al. 2008;
Sullivan et al. 2009; Foley et al. 2012), since the rest-frame
spectrum in the UV shifts to that in the optical region due to cosmic
expansion. All of these datasets suggest significant dispersion in the
UV for SNe~Ia at both low and high redshifts. However, precise
evolutionary constraints are still limited by the absence of
high-quality, low-redshift UV spectral data.

In {\it HST} Cycle 13 (year 2004--2005), extensive UV observations of
low-redshift SNe~Ia were conducted (program GO-10182; PI Filippenko).
Unfortunately, with the failure of STIS shortly before the project
started, the PR200L prism became the only available element on {\it
  HST} capable of UV spectroscopic observations, yielding only
low-resolution but still useful UV spectra of four SNe~Ia (SNe 2004dt,
2004ef, 2005M, and 2005cf). These four objects represent three
different subclasses of SNe~Ia. Amongst our sample, SN 2004dt, and
perhaps SN 2004ef, can be put into the high-velocity (W09a) or the
high-velocity-gradient groups (Benetti et al. 2005).  Moreover, SN
2004dt has the highest line polarization ever observed in a SN~Ia
(Wang et al. 2006), and is a clear outlier in the relationship between
the Si~II velocity gradient and the nebular-phase emission-line
velocity shift (Maeda et al. 2010). SN 2005M is a spectroscopically
peculiar object like SN 1991T (Filippenko et al. 1992a; Phillips et
al. 1992), with the early spectrum showing weak Ca II H\&K lines and
prominent Fe~III lines (Thomas et al. 2005). And SN 2005cf has been
regarded as a standard SN~Ia (e.g., Wang et al. 2009b, hereafter
W09b), belonging to the ``Normal'' or the low-velocity-gradient
group. Our {\it HST} UV observations of the above four events reveal
significant diversity of the UV properties among SNe~Ia. Good
understanding of the UV diversity forms the basis for further
improvements in the application of SNe~Ia as cosmological probes.

This paper is structured as follows. Observations and data reduction
are described in \S 2, while \S 3 presents our results and a
comparison with the existing UV data. Discussions are given in \S 4,
and we summarize our conclusions in \S 5.

\begin{figure}[htbp]
\includegraphics[angle=0,width=85mm]{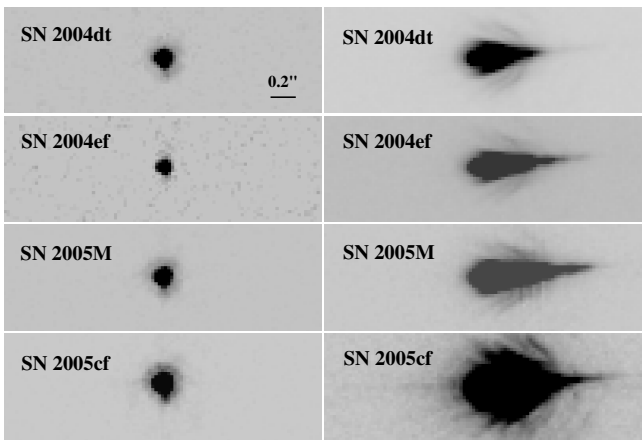}
\vspace{0.2cm}
\caption{Left panels: Direct images (F330W) of SNe 2004dt, 2004ef,
  2005M, and 2005cf near maximum light obtained with {\it HST}
  ACS/HRC.  Right panels: Raw images near maximum light obtained in
  the ACS/HRC PR200L slitless spectroscopy mode. The dispersed image is only
  a few times longer than the point-spread function of the direct images,
  indicating that the {\it HST} prism spectra have very low
  resolution.}
\label{fig-1} \vspace{-0.0cm}
\end{figure}

\section{Observations and Data Reduction}

UV observations of SNe 2004dt, 2004ef, 2005M, and 2005cf were
performed using the Advanced Camera for Surveys (ACS) high-resolution
channel (HRC) in both the slitless prism spectroscopy and the direct
imaging modes, as shown in Figure 1. Observations with the HRC PR200L
generally consist of a ``direct'' or undispersed image taken through
the F330W filter, used to establish the zero point of the wavelength
scale, and followed by dispersed images with one or more exposures
through the PR200L prism. The parallel imaging observations were made
through the F220W, F250W, and F330W filters, with the exposure time
split into several segments. Individual exposures were combined using
the MultiDrizzle task within the Space Telescope Science Data Analysis
System (STSDAS) to reject cosmic rays and perform geometric-distortion
corrections. A log of the observations is given in Table 1, listing
the exposure times of the prism and direct imaging exposures.

The UV spectra were extracted in PyRAF using the ``aXe'' slitless
spectroscopy reduction package (K\"{u}mmel et al. 2004, 2009). The
spectral extraction within aXe relies upon the position, morphology,
and photometry of the targets as observed in the accompanying direct
image. Note that the dispersion of the PR200L prism varies with
wavelength in a highly non-linear fashion. For example, the spectral
resolution is 5~\AA\ pixel$^{-1}$ at 1800~\AA\, and
105~\AA\ pixel$^{-1}$ at 3500~\AA, and it decreases rapidly to
$\sim$563~\AA\ pixel$^{-1}$ by 5000~\AA; see Table 4 of Larsen
et~al. (2006) for a detailed dispersion wavelength relation for the
PR200L prism data. The rapidly decreasing dispersion toward longer
wavelengths results in an extreme ``red pile-up'' effect for the
PR200L data (with flux at wavelengths longer than 4000~\AA\ being
compressed into just a few pixels), making studies of intrinsically
red objects problematic (see discussion in \S3.2). To decrease the
contamination of red light toward shorter wavelengths, we chose a
proper extraction window by taking advantage of the fact that the
diffraction spikes from the red pile-up are at an angle to the
dispersion.

In the reduction of the direct images, we performed aperture
photometry on the drizzled images using an aperture radius of 4 pixels
($\sim 0\farcs1$) to get the optimal signal-to-noise ratio (S/N).
Inspection of the {\it HST} ACS images taken with the HRC (with a
spatial resolution $\sim 0\farcs025$ pixel$^{-1}$) shows that there is
no significant emission from the diffuse source (see Fig. 1); thus, we
assume that the contamination of the host galaxy can be neglected for
our UV photometry. The sky background level was determined from the
median counts in an annulus of radius of 100--130 pixels. The measured
magnitudes were further corrected to an infinite-radius aperture and
placed on the Vega magnitude system (Sirianni et al. 2005). The
aperture corrections from $\sim0\farcs1$ to an infinite aperture are
0.41 mag, 0.38 mag, and 0.35 mag for the F220W, F250W, and F330W
filters, respectively. The photometric calibration of ACS/HRC was
obtained with the observations of several spectrophotometric standards
and has an uncertainty better than 2\% (Pavlovsky et~al. 2004), which
we adopt as the uncertainty for our UV photometry.

Sources of error in obtaining absolute photometry include
uncertainties in the red-leak correction and the $K$-correction. The
red-leak effect, a common feature of UV filters, is contamination from
an off-band flux in the red part of the spectrum. For stars with known
colors or spectral types, this effect can be accurately quantified
(Pavlovsky et~al. 2004; Chiaberge \& Sirianni 2007; Brown et
al. 2010). The red-leak correction is negligible for filter F330W, and
is very small for filter F250W (e.g., $\lesssim$2\% for stars with
colors bluer than the G type). This correction is greater for filter
F220W because of a significant off-band flux contamination in this
band,\footnote{Following Chiaberge \& Sirianni (2007), the cutoff
  wavelength between in-band and off-band is chosen as a filter's 1\%
  transmission point, which occurs at 2860~\AA, 3420~\AA, and
  3710~\AA\ for F220W, F250W, and F330W, respectively.}, ranging from
$\sim 2$\% for an A-type star (corresponding to a SN~Ia at early
phases) to $\sim 12$\% for a G-type star (corresponding to a SN~Ia in
the nebular phase). Despite the fact that all of our SNe~Ia have low
redshifts, $K$-corrections (Oke \& Sandage 1968) are non-negligible in
the UV due to a large variation in the spectral flux at shorter
wavelengths. In calculating the $K$- and red-leak corrections for the
UV photometry, we have used the corresponding PR200L prism spectrum
mangled to match with the observed photometry (see \S 3.2) or a
template from Hsiao et~al. (2007) when the prism spectrum has
extremely low quality or is not available. Uncertainties in the above
two corrections can be derived from the fluctuations of the input
spectra, which are determined by the photometric errors (for the
mangled prism spectra) or are taken to be $\sim 10$\% of the flux (for
the template spectrum from Hsiao et al.). The final {\it HST} ACS UV
magnitudes of SNe 2004dt, 2004ef, and 2005M, as well as the
corresponding $K$- and red-leak corrections, are listed in Table 2;
corresponding data for SN 2005cf have been published by W09b.

The distances to the SNe were computed by using the recession velocity
$v$ of their host galaxies (see Table 3), with a Hubble constant
$H_{0} = 74$ km s$^{-1}$ Mpc$^{-1}$ (Riess et al. 2009).  For objects
in the Hubble flow ($v_{\rm helio} \gtrsim 3000$ km s$^{-1}$) (SNe
2004dt, 2004ef, and 2005M), the recession velocity was determined with
respect to the 3~K cosmic microwave background radiation; for the
nearby sample with $v_{\rm helio} < 3000$ km s$^{-1}$ (SN 2005cf), the
velocity was corrected to a 220 km s$^{-1}$ Virgocentric infall.

\section{Results}
\begin{figure}[htbp]
\figurenum{2}
\hspace{-0.6cm}
\vspace{-8.5cm}
\includegraphics[angle=0,width=110mm]{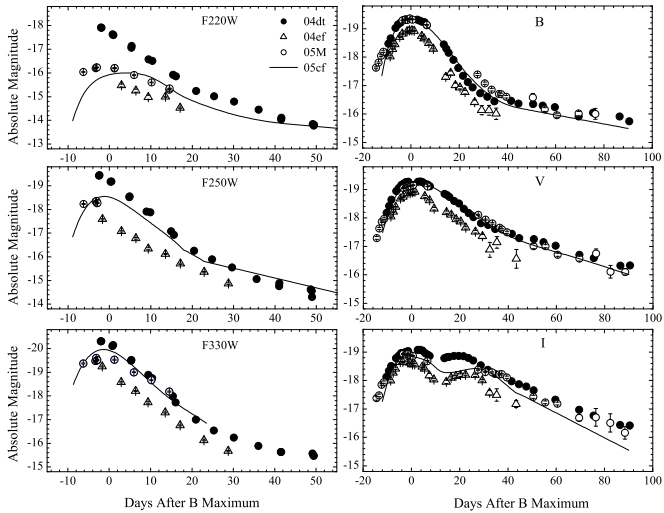}
\vspace{8.0cm}
\caption{UV light curves of the Hubble-flow SNe 2004dt, 2004ef, and 2005M
  obtained with the {\it HST} ACS/HRC and the F220W, F250W, and F330W
  filters, together with the corresponding optical light curves in the
  $BVI$ bands (Ganeshalingam et al. 2010). Overlaid are the UV-optical
  template light curves of the ``gold standard'' SN~Ia 2005cf (W09b).}
\label{fig-2} \vspace{-0.0cm}
\end{figure}

\subsection{Ultraviolet Light Curves}

Figure 2 shows the UV light curves of SNe 2004dt, 2004ef, and 2005M
obtained with the {\it HST} ACS and the F220W, F250W, and F330W
filters. Also plotted are the corresponding optical light curves (in
$BVI$) from Ganeshalingam et al. (2010). As SN 2005cf was well observed
and is judged to be a normal SN~Ia, and its light curves have been
published elsewhere (W09b), we use it as a template SN~Ia for comparison study.
To account for effects of the redshift on the observed SN flux, we applied
$K$-corrections (Oke \& Sandage 1968) to all of the light curves using
the observed spectra (as shown in Figures 5 and 6) and the Hsiao et
al. (2007) template.

The light curves have also been corrected for reddening in the host
galaxy as well as in the Milky Way. The total reddening toward a SN~Ia
can be derived from an empirical relation established between decline
rate $\Delta m_{15} (B)$ and color indexes for a low-reddening sample
(e.g., W09b, and references therein). By comparing the observed color
with that predicted by the $\Delta m_{15}(B)$ vs. $(B_{\rm max} -
V_{\rm max}$) relation, we estimate the total reddening $E(B - V)_{\rm
  total}$ as 0.10 mag for SN 2004dt, 0.17 for SN 2004ef, 0.12 mag for
SN 2005M, and 0.20 mag for SN 2005cf, with a typical error of 0.03
mag.  The absorption in the Milky Way (MW) was estimated by using the
reddening maps of Schlegel, Finkbeiner, \& Davis (1998) and the
Cardelli, Clayton, \& Mathis (1989) reddening law with $R_{V} = 3.1$,
while extinction in the host galaxy was corrected applying an
extinction law with $R_{V} \approx 2.3$ (e.g., Riess et al. 1996;
Reindl et al. 2005; Wang X. et al. 2006; Kessler et al. 2009). Since all
four SNe~Ia have low host-galaxy reddening (e.g., $E(B - V)_{\rm host}
\lesssim 0.10$ mag), a moderate change of the extinction-law
coefficient will have limited effect on the UV light curves shown in
Figure 2. According to the extinction-law coefficients recalculated by
Brown et~al. (2010), a change in $R_{V}$ from 1.8 (LMC dust) to 3.1
(Milky Way dust) will only result in a difference of $\sim 0.02$ mag
in F330W and $\lesssim 0.15$ mag in F250W for a SN~Ia with a
host-galaxy reddening $E(B - V)_{\rm host} = 0.10$ mag.  Note,
however, that the magnitude change in F220W could be larger due to a
dramatic variation in the resultant extinction coefficient $R_{\rm
  F220W}$. Table 3 lists the relevant parameters of these four SNe~Ia.

\begin{figure}[htbp]
\figurenum{3}
\vspace{-1.0cm}
\hspace{-0.5cm}
\includegraphics[angle=0,width=95mm]{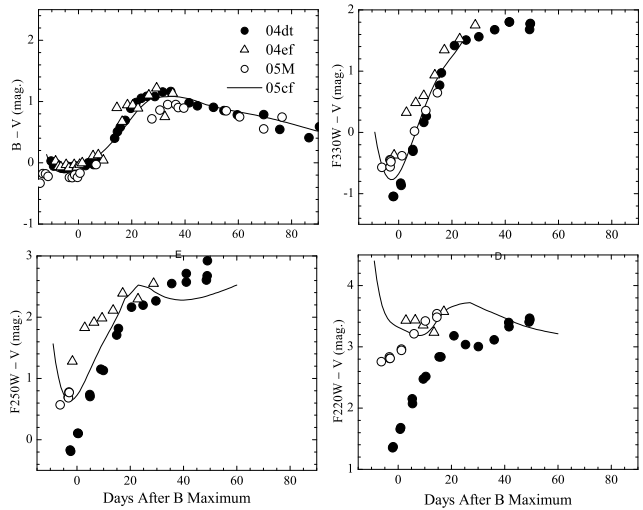}
\vspace{-0.6cm}
\caption{The UV color curves of the Hubble-flow SNe 2004dt, 2004ef, and
  2005M. Overlaid are the color-curve templates of the ``gold
  standard'' SN~Ia 2005cf (W09b).}
\label{fig-3} \vspace{-0.0cm}
\end{figure}

One can see from Figure 2 that the UV light curves of our SNe~Ia
show a large dispersion, especially at shorter wavelengths, while the
optical counterparts are generally similar except in the $I$ band
where there is more diversity. Of our sample, SN 2004ef appears
relatively faint in all bands, consistent with the fact that it is a
slightly faster decliner with $\Delta m_{15}(B) \approx 1.46 \pm 0.06$
mag. SN 2005M is a spectroscopically peculiar object like SN 1991T. This
supernova has a slower decline rate ($\Delta m_{15}(B) = 0.86 \pm 0.05$ mag),
but it does not appear to be overluminous either in the UV or optical bands.
However, our relative ignorance regarding the intrinsic color of these
slow decliners might prevent us from deriving reliable reddenings for them.
A notable feature in the plot is that the UV emission of SN 2004dt appears
unusually strong around $B$-band maximum and decreases rapidly after maximum
light. Moreover, SN 2005M seems to peak a few days earlier than SN
2005cf in the F220W and F250W bands. Such a temporal shift seems to
exist for SN 2004dt and SN 2004ef, although their light curves in UV
are not well sampled around maximum brightness.

The color curves of our four SNe~Ia, corrected for the MW and
host-galaxy reddening, are presented in Figure 3. One can clearly see
that their ${\rm UV} - V$ curves show remarkable differences despite having
similar $B-V$ curves. SN 2004dt is quite peculiar, being very blue
compared with the other three objects. It is worthwhile to point out
that the {\it Swift} UVOT has seen a number of events (~7) that are also blue
in the NUV-optical photometric colors (private communication with P. Milne).

Since SN 2004dt has $\Delta m_{15}(B) \approx 1.1$ mag and a peak
$B-V$ color ($\sim 0.0$ mag) similar to that of SN 2005cf, it is of
interest to perform a detailed comparison of their UV properties.
Around maximum light, SN 2004dt is bluer than SN 2005cf by $\sim 0.8$
mag in ${\rm F250W} - V$ and by $\sim 2.0$ mag in ${\rm F220W} - V$. About 2--3
weeks after maximum, the color difference becomes less significant as
a result of the rapid decline of the post-maximum UV emission in SN
2004dt (see Fig. 6 and the discussion in \S 3.2). In the first 15 days
after the maximum, the decline of the UV light curve is measured to be
$\sim 1.9$ mag in F220W, $\sim 2.4$ mag in F250W, and $\sim 2.5$ mag
in F330W for SN 2004dt. These are all larger than the corresponding
values measured for SN 2005cf by $\sim 0.6$--0.9 mag (see also Table
10 in W09b). A faster post-maximum decline usually corresponds to a
shorter rise time for the light curves of SNe~Ia in the optical bands,
as evidenced by the fact that they can be better normalized through
the stretch factor (Perlmutter et al. 1997; Goldhaber et
al. 2001). For SN 2004dt, a shorter rise time in the UV is consistent
with the high expansion velocity that could make the gamma-ray heating
become less effective. Moreover, a higher expansion velocity could
extend the radius of the UV photosphere to higher velocities and hence
result in a larger UV flux. The possible origin of the UV excess in SN
2004dt in the early phase is an interesting issue and is further
discussed in \S 4.2.

\subsection{Ultraviolet Spectroscopy}

As shown in Table 1, a total of 34 {\it HST} ACS PR200L prism spectra
have been collected for SNe 2004dt, 2004ef, 2005M, and 2005cf. In
column (5) of Table 1, we list the total exposure time for each
spectrum, resulting from a series of coadded CR-split exposures. Among
the four SNe~Ia observed with the {\it HST} ACS, SN 2005cf and SN
2004dt have better temporal sequences and higher S/N for their UV
spectra (see Fig. 5 and 6). Owing to very low spectral resolution at
longer wavelengths, our analysis in this paper is restricted to the
useful wavelength range 1800--3500~\AA.

To evaluate the quality of the prism data, we compare the synthetic
magnitudes derived from the UV spectra with those obtained by direct
imaging observations for SN 2005cf and SN 2004dt (see Figure 4). In
F330W, these two measurements match well with each other near maximum
light, but the spectrophotometric magnitudes decline more slowly
starting about one week after maximum. A similar trend in the decline
rate exists in the F220W and F250W bands of SN 2004dt, and in the
F250W band of SN 2005cf. However, compared with the corresponding
imaging observations, it is alarming that the flux of the prism
spectra is too high in the F250W band ($\sim$ 0.6 mag for SN 2004dt, $\sim$0.7 mag
for SN 2005cf), and especially in the F220W band ($\sim$1.0 mag for SN 2004dt,
$\sim$2.0 mag for SN 2005cf). This suggests that the flux level of PR200L calibrated
by the white dwarfs (which are very blue) may not be accurate for the flux calibration
of the supernova.

According to ACS/HRC Instrument science report ACS 2006-03 (Larsen
et~al. 2006), the flux calibration of the PR200L prism spectra based
on white dwarfs is expected to be accurate to $\sim 5$\% over the
wavelength range 1800--3000~\AA. However, red objects usually suffer
systematic errors in the UV flux calibration because the blue end of
the spectrum can be highly contaminated by scattered light from the
red end as a result of the ``red pile-up'' effect. Such an effect is
probably a dominant factor accounting for the higher F250W and
F220W fluxes, as indicated by the wavelength-dependent discrepancy between
the spectrophotometry and observed photometry\footnote{One can also see
from Fig.1 that this effect is particularly obvious for SN 2005cf
as a lot of photons have spilled over from the red to the blue of the dispersed image},
as well as for the slower post-maximum decline seen in all three UV bands for the
SNe in our sample; SNe~Ia are known to become progressively redder
after maximum light and the contamination of the red light becomes
increasingly significant. However, the red pile-up effect is at
present poorly quantified, and more test observations of objects
covering a wide range of colors are clearly needed to better calibrate
PR200L prism spectra.

\begin{figure}[htbp]
\figurenum{4}
\vspace{-0.5cm}
\hspace{0.5cm}
\includegraphics[angle=0,width=80mm]{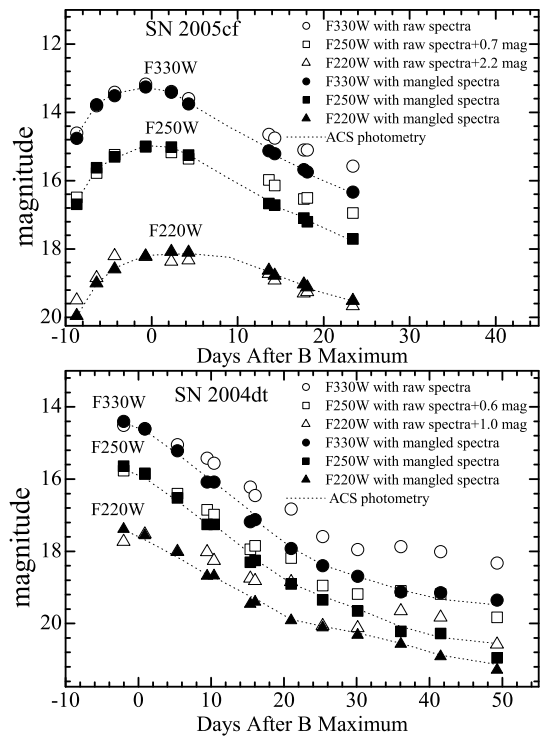}
\vspace{-0.5cm} \caption{Comparison of the spectrophotometry with the
  observed photometry for SN 2004dt and SN 2005cf at UV
  wavelengths. The open symbols represent the synthetic magnitudes
  from the raw spectra, and the filled symbols represent the values
  from the spectra mangled to the observed photometry. The dashed
  curves represent the observed photometry.}
\label{fig-4} \vspace{-0.2cm}
\end{figure}

Fortunately, we also have ACS HRC photometry taken nearly at the same
time, making it possible to obtain accurate relative flux calibration
of the PR200L prism spectra. Given the relatively smooth variations in
the raw prism spectra, it is not difficult to model the continuum of
the spectra with a simple polynomial function.  The final calibrated
spectra of SN 2005cf and SN 2004dt are shown in Figures 5 and 6; the
spectrophotometry is consistent with the photometric results within
$\pm$0.05 mag. However, the observed increase in flux below about
2200~\AA\ is almost certainly an artifact and should not be trusted.

The UV spectra of SN 2004ef have poor S/N; the object is relatively
faint, being at a larger distance ($\sim 120$ Mpc). Only a single
UV-prism spectrum was obtained of SN 2005M. The UV spectra of SN
2004ef and SN 2005M are shown in Figure 7.

\subsubsection{SN 2005cf}

Figure 5 shows the evolution of the combined UV-optical spectrum of SN
2005cf from $t = -9$ days to $t = +24$ days relative to the $B$-band
maximum. The ground-based spectra are taken from W09a and Garavini et
al. (2007). Spectra obtained within 1--2 days of each other were
combined directly. In cases where no contemporaneous optical spectrum
was available, an interpolated spectrum was used. After flux
calibration with the corresponding ACS/HRC photometry, the UV-prism spectra
agree reasonably well with their optical counterparts over the
spectral range 3300--3500~\AA\ (see Figure 5).

\begin{figure}[htbp]
\figurenum{5}
\vspace{-0.5cm}
\hspace{-0.5cm}
\includegraphics[angle=0,width=90mm]{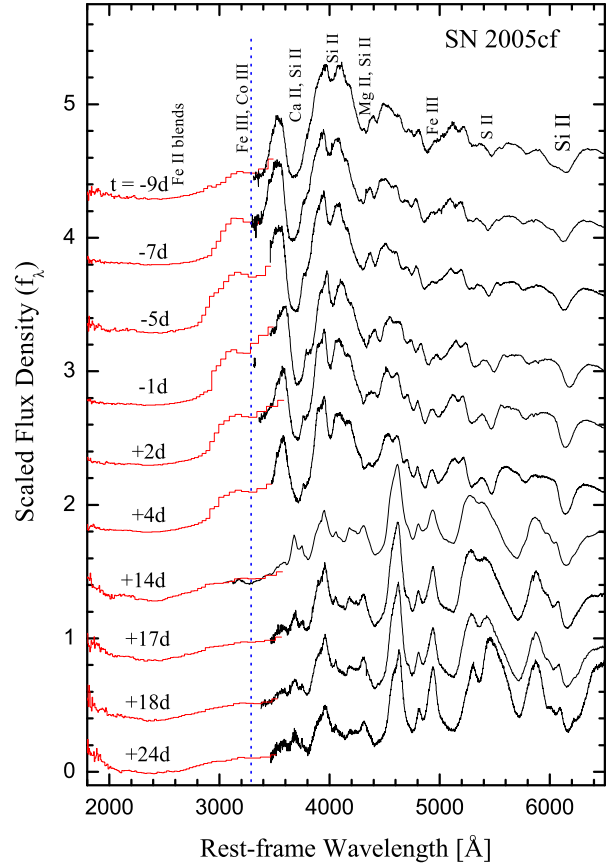}
\vspace{-0.5cm} \caption{Evolution of the UV-optical spectrum
  of SN 2005cf. The UV spectra were obtained with the {\it HST} ACS prism
  (PR200L), and the corresponding optical data are taken from W09b and
  Garavini et al. (2007). All of the spectra have been rescaled to
  match the UV-optical photometry, and were adjusted to the peak of the
  spectral flux and arbitrarily shifted for display. The dashed line marks
  the position of the $\sim$ 3250 \AA\ absorption feature. The
  increase in flux below about 2200~\AA\ is almost certainly an
  artifact and should not be trusted.}
\label{fig-5} \vspace{-0.2cm}
\end{figure}

It is known that in the UV spectra of SNe~Ia, almost no individual
lines can be identified. This is because the vast number of iron-group
element lines in the UV overlap strongly, eliminating individual line
features. This contrasts with the optical, where P-Cygni lines of
neutral or singly ionized ions of intermediate-mass elements (mostly
O, Mg, Si, S, and Ca) can usually be identified. Probably the only
individual feature in SN~Ia UV spectra blueward of $\sim
3700$~\AA\ that can be obviously identified is the P-Cygni line
(absorption feature at $\sim 2650$~\AA) caused by Mg~II $\lambda$2800
which was seen in {\it HST} spectra of SN 1992A at $t=5$ days after
$B$ maximum (Kirshner et al. 1993). In our spectra, we see no clear
evidence of this line or any other line-like feature blueward of $\sim
2800$~\AA. In most of our spectra, this is probably caused by the low
S/N and low resolution.  In the case of SN~2004dt, the effect that led
to strong UV emission (e.g., circumstellar matter interaction) may
have smoothed out line-like features (see the discussion below).

Inspecting the near-maximum UV spectra in the wavelength range
1800--3500~\AA, we identify a prominent absorption feature near
3250~\AA. It is also seen in the {\it Swift} UVOT UV-grism spectrum of
SN 2005cf near maximum light (e.g., Bufano et al. 2009; see also
Fig. 9) and the STIS UV spectra of the Palomar Transient Factory
sample of SNe~Ia (Cooke et al. 2011). However, we note that the
relatively weak absorption feature near $\sim 3050$~\AA\ seen in STIS
and {\it Swift} spectra is nearly invisible in the {\it HST} UV
spectrum, indicating the limited resolution of the {\it HST}/ACS
prism. The resolution effect can be demonstrated by degrading the
{\it Swift} spectrum. The $\sim 3050$~\AA\ feature becomes very weak
in the {\it Swift} spectrum degraded to a resolution of $\sim
50$~\AA\ pixel$^{-1}$ (see the dash-dotted line in Fig. 9), and this
feature almost disappears when further decreasing the spectral
resolution to $\sim 100$~\AA\ pixel$^{-1}$ (see the dashed line in
Fig. 9). The absorption feature near 3250~\AA\ also becomes less
significant at a resolution $\sim 100$~\AA\ pixel$^{-1}$, consistent
with that observed in the {\it HST} prism spectrum.

The flux ratio $R_{UV}$, defined as
$f_{\lambda}$(2770~\AA)/$f_{\lambda}$(2900~\AA), has been proposed to
correlate with the peak luminosity of SNe~Ia (Foley, Filippenko, \&
Jha 2008). Our {\it HST} spectra are not appropriate for testing
such a correlation due to their limited resolution ($\gtrsim
40$~\AA\ pixel$^{-1}$ at 2800--2900~\AA), which could have significant
impact on the intrinsic UV slope within such a narrow wavelength
range. Even more importantly, as discussed in \S 3.2, red pile-up
affects the spectra substantially in this region.

The near-UV features in the $\sim 2700$--3500~\AA\ range of the
early-time spectra of SNe~Ia were initially studied by Branch \&
Venkatakrishna (1986), who suggested that they are produced by blends
of Fe~II and Co~II lines. In particular, the $\sim
3250$~\AA\ absorption feature has been attributed to blueshifted Co~II
absorption (rest wavelength 3350--3500~\AA; Branch et al. 1985), and
the absorption feature at $\sim 3050$~\AA\ is ascribed to Fe~II
absorption. Based on their study of the prominent example SN 1992A,
Kirshner et al. (1993) confirmed that the near-UV region is largely
formed by a complex blend of iron-group element lines. They found that
Cr~II, Mn~II, and Fe~II contribute significantly to the
absorption-like feature at $\sim 3250$~\AA, while the ions Co~II and
Ni~II make no significant contribution because the newly synthesized
Ni-Co is confined to the inner regions at these early phases. The
spectral properties of SNe~Ia in the UV were also investigated by
Sauer et al. (2008), based on the early-time spectra of SN 2001eh and
SN 2001ep obtained with {\it HST} STIS. In their study, the
contribution of Ti~II to the $\sim 3050$~\AA\ feature and of
doubly ionized species such as Fe~III and Co~III to the $\sim
3250$~\AA\ features of SN 2001eh are important.  The absorption
features blueward of 2700~\AA\ in the spectrum are also thought to
originate from singly and doubly ionized Fe and Co (Kirshner et
al. 1993; Sauer et al. 2008).

\begin{figure}
\figurenum{6}
\vspace{-0.5cm}
\hspace{-0.5cm}
\includegraphics[angle=0,width=90mm]{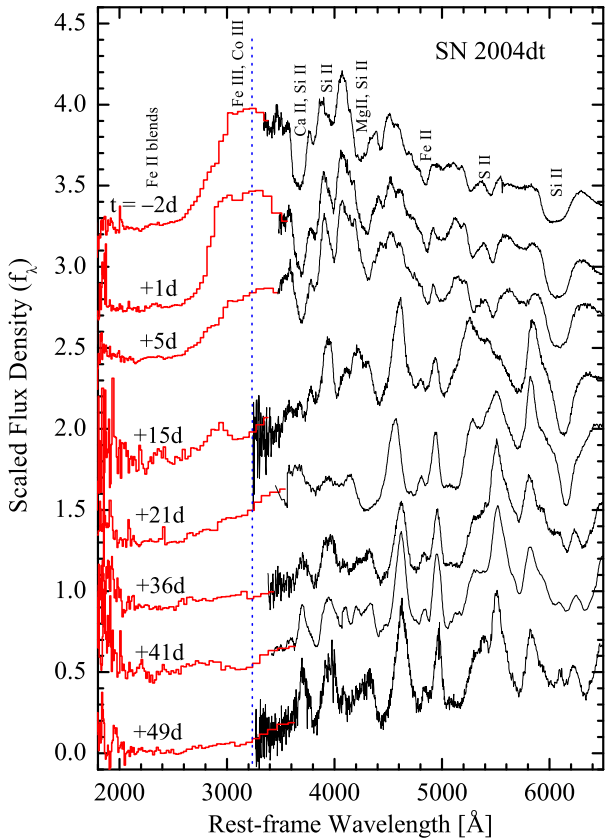}
\vspace{-0.5cm}
\caption{Evolution of the UV-optical spectrum of SN 2004dt. The UV
  spectra were obtained with the {\it HST} ACS/HRC PR200L prism, and
  the optical data are from Altavilla et al. (2007).  All of the
  spectra have been rescaled to match the UV-optical photometry, and
  were normalized to the flux peak of the spectrum and arbitrarily shifted
  for display. The vertical dashed line marks the position of the
  absorption feature near $\sim$ 3250 \AA\ .}
\label{fig-6}
\end{figure}

\subsubsection{SN 2004dt}

The evolution of the combined UV-optical spectrum of SN 2004dt is
displayed in Figure 6. The UV spectroscopic observations of SN 2004dt
cover the period between $t \approx -2$ days to $t \approx +50$ days
with respect to the $B$-band maximum. The optical data shown in the
plot are taken from Altavilla et al. (2007). At $t = -2$ days and $t =
+1$ days, it is evident that there is a prominent peak emerging
between 2900~\AA\ and 3400~\AA. The absorption-like feature at $\sim
3250$~\AA, as seen in SN 2005cf and other normal SNe~Ia (Bufano et
al. 2009, and this paper), is very weak and nearly invisible in SN
2004dt. This feature becomes prominent in the $t = +15$ day and $t =
+41$ day spectra, but it is weak at other epochs after maximum
brightness. This variation may be caused by the poor S/N, the lower
resolution of the PR200L spectrum, possible errors in the data
reduction, or a combination of the above factors. Moreover, the
overall spectral flux in the UV region drops rapidly after maximum
light, consistent with the rapid decline of the UV light curves (see
Fig. 2).

The strengthening of the $\sim 3200$~\AA\ feature in the post-maximum
spectra is particularly interesting in the case of SN 2004dt. It may
reflect an increase in the line blanketing opacity as a result of a
retreat of the photosphere into the iron-group-rich inner region of
the ejecta, which is supported by the fact that the $\sim
5000$~\AA\ feature (due to blends of Fe~II and Fe~III) gradually
gained strength after maximum brightness. The Doppler velocity of
Si~II $\lambda$6355 is found to be $\sim$ 14,000 km s$^{-1}$ for SN
2004dt from the near-maximum spectrum, which fits comfortably in the
high-velocity category (W09a). This object is also a member of the
high-velocity-gradient group (Benetti et al. 2005) and the broad-line
(BL) subclass (Branch et al. 2006). It is interesting to examine
whether the distinct UV behavior correlates with the production of the
high-velocity features in SNe~Ia. The origin of the high-velocity
features is still unclear. They could be due to an increase of the
abundance and/or density in the outer layers, circumstellar material
interaction (e.g., Gerardy et al. 2004; Mazzali et~al. 2005), or a
clumpy distribution of the outer ejecta (Leonard et al. 2005;
Wang et al. 2006).

\begin{figure}
\figurenum{7}
\vspace{-0.5cm}
\hspace{-0.5cm}
\includegraphics[angle=0,width=95mm]{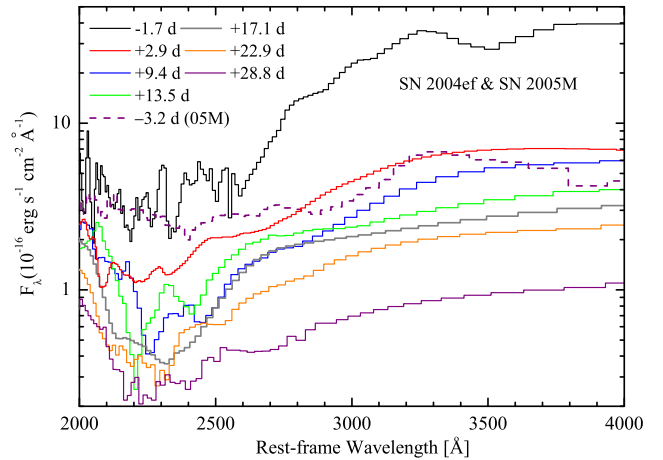}
\vspace{-0.8cm}
\caption{Evolution of the UV spectra of SN 2004ef (solid curves) and
  SN 2005M (the dashed curve).  The UV spectra were obtained with the
  {\it HST} ACS/HRC PR200L prism. The increase in flux below about
  2200~\AA\ is almost certainly an artifact and should not be
  trusted.}
\label{fig-7}
\end{figure}

\subsubsection{SN 2004ef and SN 2005M}

In Figure 7, we show the evolution of the UV spectrum of SN 2004ef, as
well as a single-epoch spectrum of SN 2005M.  The UV spectroscopic
observations of SN 2004ef span from $t \approx -2$ days to $t \approx
+29$ days with respect to the $B$-band maximum. A temporal sequence of
the combined UV-optical spectra cannot be constructed for SN 2004ef
due to the sparse optical data at similar phases.

As in SN 2005cf and SN 2004dt, the UV spectra of SN 2004ef show a
broad and shallow absorption trough near 2300~\AA, which is likely due
to blending of the multiplets of the ionized iron-group elements
(Kirshner et~al. 1993). Note that the prominent absorption feature near
3250~\AA, commonly seen in other SNe~Ia, is invisible in the UV
spectra of SNe 2004ef and 2005M. This inconsistency reminds us that
the limited S/N and low resolution of the spectra could have
significant influence on the observed features of the spectra.

\begin{figure*}[!ht]
\figurenum{8}
\hspace{1.5cm}
\vspace{-2.5cm}
\includegraphics[angle=0,width=150mm]{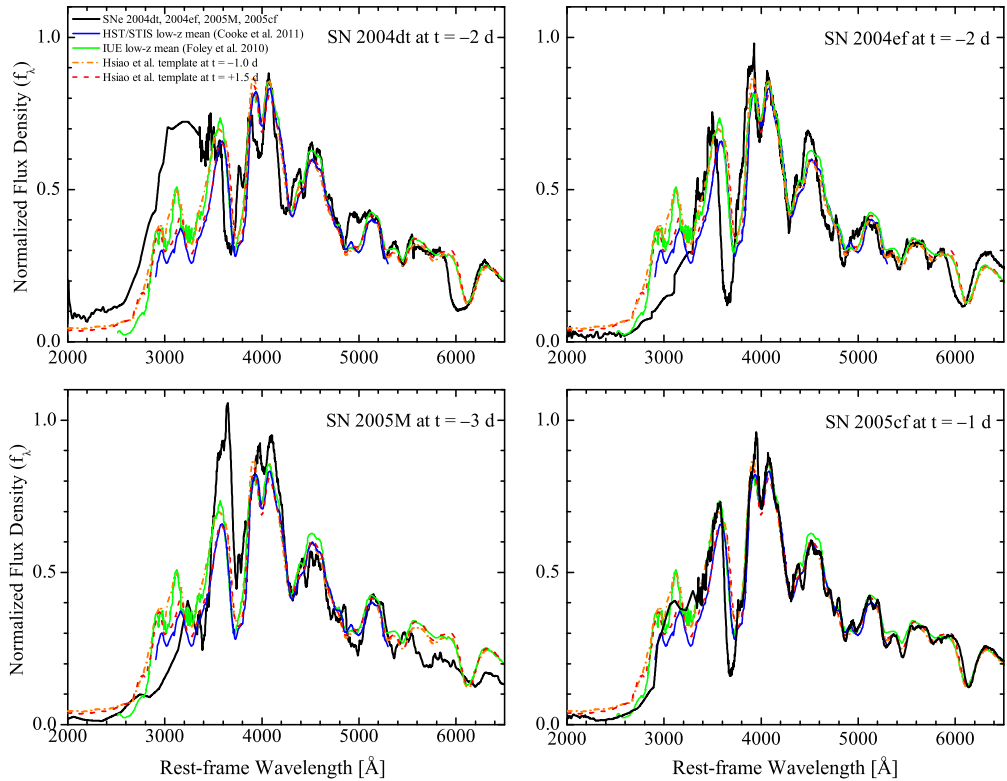}
\vspace{2.0cm}
\caption{Comparison of the near-maximum-light UV-optical spectra of
  SNe 2004dt, 2004ef, 2005M, and 2005cf with the mean spectra of the
  low-redshift SNe~Ia constructed by Cooke et al. (2011), Foley et
  al. (2012), and Hsiao et al. (2007). Note that the low resolution of
  our {\it HST} spectra may play a substantial role in affecting the
  spectral features below 3500~\AA.}
\label{fig-8}
\end{figure*}

\subsubsection{Comparison of the UV spectra}

In Figure~8, we compare the near-maximum UV spectra of our four nearby
SNe~Ia with other mean low-redshift SN~Ia spectra at similar
phases. The spectra have been corrected for reddening in the Milky
Way. The mean low-redshift spectra come from several studies,
including the mean spectrum constructed primarily with the {\it HST}
STIS spectra (Cooke et al. 2011), the combined spectrum of a few
nearby SNe~Ia with archival {\it HST} and {\it IUE} data (Foley et
al. 2012), and the spectral template established by Hsiao et
al. (2007).  As Cooke et al. (2011) did not apply any extinction
correction in building their mean low-redshift spectrum, we apply a
reddening correction to their spectrum, with $E(B-V) = 0.10$ mag and
$R_{V} = 3.1$, to correct for the color difference.  All of the
spectra shown in the plot are normalized to approximately the same
flux in the rest-frame wavelength range 4000--5500~\AA. Note that the
Hsiao et al. (2007) template spectra contain high-redshift SN~Ia data
from the Supernova Legacy Survey (Ellis et al. 2008); hence, they may
suffer from evolutionary effects and may not represent the genuine
spectrum of local SNe~Ia in the UV region.

\begin{figure}[htbp]
\figurenum{9}
\vspace{0.0cm}
\hspace{-0.8cm}
\vspace{-0.2cm}
\includegraphics[angle=0,width=100mm]{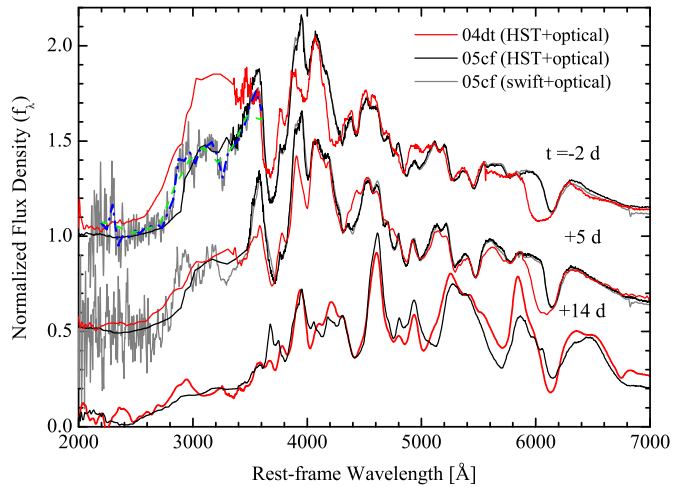}
\caption{Comparison of the UV-optical spectra of SN 2004dt and SN
  2005cf at three selected epochs ($t = -2$, +5, and +14 with respect
  to $B$-band maximum). The dash-dotted (blue) and dashed (green)
  lines represent the {\it Swift} UV spectra of SN 2005cf degraded to
  a resolution of $\sim 50$~\AA\ and $\sim 100$~\AA, respectively.}
\label{fig-9}
\end{figure}

The Foley et al. spectrum (mean phase $t \approx 0.4$ days) and the
Hsiao et al. $t = -1$ day spectrum agree fairly well within the
uncertainties, and the STIS UV spectrum (mean phase $t \approx 1.5$
days) also agrees shows reasonable agreement given the difference in
epoch.  We see that there is a general shift of UV peak features to
the red between $t \approx -1$ day and $t \approx +1$ day. Most
notably, the large peak near 3100~\AA\ at $t \approx -1$ day shifts to
$\sim 3150$~\AA\ at $t \approx +1$ day. The underlying cause of the
shift is probably just the recession in velocity of the UV photosphere
with the expansion of the ejecta, but detailed modeling would be
needed to verify this.

Of our sample, SN 2004dt appears to be unusually bright in the UV. Its
UV emission is found to be stronger than the Foley et al. low-redshift
mean spectrum for $t \approx -1$ day by $\gtrsim 75$\% (or at a
confidence level $\gtrsim 6$--7$\sigma$) at wavelengths
2500--3500~\AA.  Differences are also present in the optical portion
of the spectrum where the absorption features of intermediate-mass
elements (IMEs) are relatively strong and highly blueshifted (except
for the S~II lines), although the integrated flux over this region
does not show significant differences with respect to the
templates. The $\sim 5000$~\AA\ absorption feature due to Fe~II and
Fe~III blends appears to be rather weak, perhaps suggestive of a
smaller amount of iron-peak elements in the outer ejecta of SN 2004dt.

The near-maximum spectrum of SN 2004ef displays prominent features of
both IMEs and Fe-group elements, with the UV emission being weaker
than the local composite spectrum by about 25\%. Another notable
aspect of Figure 8 is the deficiency of the UV emission in SN 2005M.
This supernova exhibits the prominent Fe~III feature near 5000~\AA\ in
the earliest spectrum (Thomas 2005). SN 2005M has a smaller $\Delta
m_{15}(B)$; however, the UV flux emitted in the wavelength region
2500--3500~\AA\ is found to be even lower than the mean value by $\sim
22$\%.  One can see from Figure 8 that significant scatter is observed
in these three events at wavelengths below $\sim 3500$~\AA. In this
wavelength region, the continuum and the spectral features were
proposed to be primarily shaped by heavy elements such as Fe and Co
(Pauldrach et al. 1996). Thus, the large scatter in our sample might
be related to variations of the abundance of Fe and Co in the outer
layers of the exploding white dwarf (H\"{o}flich et al. 1998; Lentz et
al. 2000; Sauer et al. 2008). On the other hand, the spectrum of SN
2005cf generally matches well the nearby comparison except for the
Ca~II H\&K feature, which exhibits the strongest difference in the
optical region among our four events. The variations of Ca~II
and Si~II absorptions at higher velocities suggest that additional
factors, such as asphericity or different abundances in the progenitor
white dwarf, affect the outermost layers (Tanaka et al. 2008).

Figure 9 shows a more detailed comparison of the UV-optical spectra of
SN 2004dt and SN 2005cf at three different epochs ($t \approx -2$ d,
+5 d, and +14 d). The integrated fluxes of the spectra have been
normalized over the 4000--5500~\AA\ wavelength range. These two SNe~Ia
provide us good examples to examine the generic scatter in the UV
region as they have quite similar photometric properties in the
optical bands, such as the $B-V$ color at maximum and the post-maximum
decline rate $\Delta m_{15}(B)$ (see Table 3). At $t \approx -2$ day,
the normalized flux obtained for SN 2004dt at 2500--3500~\AA\ is much
brighter than for SN 2005cf in the same wavelength region, with the
flux ratio $F_{\rm 04dt}/F_{\rm 05cf} \approx 1.9$. This flux ratio in
the UV decreases quickly to $\sim 1.2$ at $t \approx +5$ day, and it
becomes roughly 1.0 at $t \approx 2$ weeks after maximum. Neither
contamination by the background light from the host galaxy nor
uncertainty in the reddening correction is likely to account for such
a peculiar evolution of the UV flux for SN 2004dt; the influence of
background emission would become more prominent as the object dims and
SN 2004dt does not suffer significant reddening in the host galaxy
(see \S 4). Such a fast decline of the UV flux is rarely seen in the
existing sample of SNe~Ia, possibly an indication of differences in
the explosion physics or progenitor environment with respect to normal
SNe~Ia.

\begin{figure*}[!ht]
\figurenum{10}
\vspace{-1.0cm}
\hspace{1.5cm}
\includegraphics[angle=0,width=150mm]{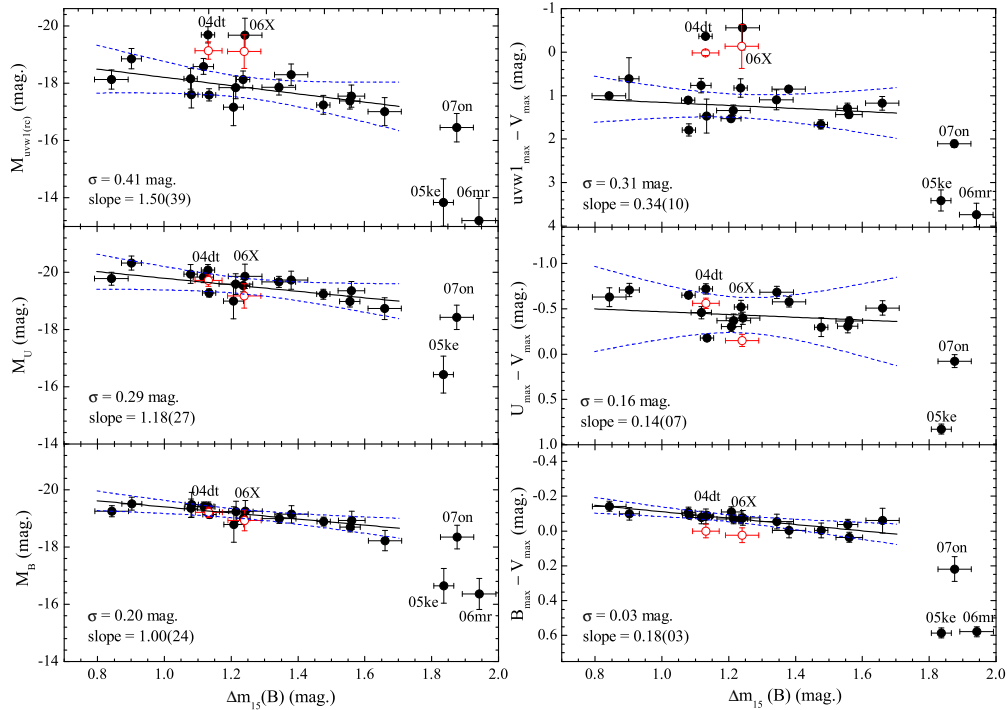}
\vspace{-1.0cm}
\caption{Left panels: the uvw1-, $U$-, and $B$-band maximum magnitudes
  versus the $B$-band decline-rate parameter $\Delta m_{15}(B)$ of
  SNe~Ia with UV observations from the {\it Swift} UVOT and {\it HST}.
  Right panels: the peak colors plotted versus $\Delta m_{15}(B)$. The
  solid lines represent the linear fit to the SNe with $0.8 < \Delta
  m_{15}(B) < 1.7$ mag, and the dashed curves represent the 3$\sigma$
  uncertainties. SN 2004dt and SN 2006X are not included in the fit,
  while the open symbols show the case assuming that these two SNe~Ia
  have intrinsically redder color by $\sim 0.1$ mag with respect to
  the other objects.}
\label{fig-10}
\end{figure*}

\subsection{General Properties of the UV \\ Luminosity of SNe Ia}

The peak luminosity of SNe~Ia in the optical bands has been
extensively studied and found to correlate well with the width of the
light curve around maximum light. Outliers have been identified as
abnormal objects with different explosion mechanisms such as the
overluminous objects SN 2003fk (Howell et~al. 2006) and SN 2009dc
(Silverman et al. 2011; Taubenberger et al. 2011), and the
underluminous events SN 1991bg (Filippenko et~al. 1992b; Leibundgut et
al. 1993) and SN 2002cx (Li et~al. 2003).  Examining the relationship
between the UV luminosity and $\Delta m_{15}(B)$ may disclose a
diversity, even for the so-called ``normal'' SNe~Ia, since the UV
emission is thought to be more sensitive to possible variations of the
explosion physics or progenitor environment.

We have examined a sample of 20 SNe~Ia with good photometry in the
UV. Four among this sample are from the {\it HST} observations
presented in this paper and the rest are {\it Swift} objects published
by Milne et al. (2010). The left panels of Figure 10 show the maximum
absolute magnitudes of these objects in broadband $U$, $B$, and {\it
  Swift} uvw1/{\it HST} F250W versus $\Delta m_{15}(B)$\footnote{As
  the instrumental response curve of the {\it HST} ACS F250W filter is
  similar to that of the {\it Swift} uvw1, the magnitudes measured in
  these two filters should be comparable. This is demonstrated by the
  observations of SN 2005cf, for which the measurements by the {\it
    HST} ACS and the {\it Swift} UVOT are consistent to within 0.1 mag
  (W09b). We thus neglect magnitude differences measured in the F250W
  and the uvw1 filters in our analysis.}. Following the analysis by
Brown et al. (2010), we applied the red-tail corrections to the
uvw1/F250W magnitudes to mitigate the effects of the optical photons
on the UV flux. The $U$ and uvw1/F250W magnitudes are also
$K$-corrected for the large variation in the spectral flux at shorter
wavelengths (but note that a UV/blue event like SN 2004dt may warrant
a second spectral sequence for the sake of red-tail and
$K$-corrections).  The peak magnitudes in the uvw1/F250W filters are
obtained by fitting the data with a polynomial or the template light
curve of SN 2005cf. The parameters in the optical bands, such as the
peak magnitudes and the $B$-band light-curve decline rate $\Delta
m_{15}(B)$, are estimated from the published light curves (see Table 3
and the references).

After corrections for the extinction (assuming an extinction law with
$R_{V} = 2.3$ for the SN host galaxy), linear fits to the subsample in
Figure 10 with $0.8 < \Delta m_{15}(B) < 1.7$ mag yield a root-mean
square scatter (i.e., $\sigma$ values) of $\sim 0.2$ mag in $B$ and
$\sim 0.4$ mag in uvw1. Note that SN 2004dt and SN 2006X were excluded
from the fit. The uvw1-band peak luminosity does show a correlation
with $\Delta m_{15}(B)$, but with steeper slopes and larger scatter
than that in $B$ and $U$. Given the fact that the dust obscuring the
SN may have different origins (e.g., Wang 2005; Goobar 2008; W09a),
the host-galaxy extinction corrections with $R_{V} = 3.1$ (Milky Way
dust) and $R_{V} = 1.8$ (LMC dust) are also applied to the absolute
magnitudes. In both cases, the scatter also increases at shorter
wavelengths. The relevant results of the best linear fit with
different values of $R_{V}$ to the $M_{\rm max} - \Delta m_{15}(B)$
relation are shown in Table 4.

The wavelength-dependent scatter can be driven in part by the
uncertainty in the absorption corrections, as the UV photons are more
scattered by the dust than the optical. Assuming a mean error $\sim
0.04$ mag in $E(B-V)_{\rm host}$ (e.g., Phillips et al. 1999; Wang et
al. 2006), the residual scatter of the uvw1-band luminosity can be as
large as $\sim 0.3$ mag, which is still much larger than that found in
$B$ ($\sim 0.15$ mag). A notable feature in the left panel of Figure 10
is the UV excess seen in SN 2004dt and perhaps SN 2006X; they are found to be brighter
than the corresponding mean value by $\sim 6.0\sigma$ and $\sim
3.0\sigma$, respectively. The discrepancy of the observed scatter
cannot be caused by the error in distance modulus, which applies
equally for the UV and optical bands. This is demonstrated by the
correlation between the peak colors and $\Delta m_{15}(B)$, as shown
in the right panels of Figure 10. The color correlation is distance
independent, but it shows significantly large scatter in the UV. In
particular, the ${\rm uvw1/F250W} - V$ colors of SN 2004dt and SN 2006X are
found to be bluer than the mean value by $\sim 1.3$ mag. This
indicates that the uvw1/F250W filter is perhaps more a probe of SN~Ia
physics than a cosmological standard candle.

Strong emission in the UV is likely an intrinsic property of SN 2004dt
rather than being due to improper corrections for the dust reddening,
since we only applied a small reddening correction to this supernova,
$E(B-V)_{\rm host} = 0.08$ mag.  Moreover, SN 2004dt still appears
much brighter and bluer in the UV than the mean values defined by the
other SNe~Ia even if it is assumed to suffer little reddening. On the
other hand, SN 2006X is heavily extinguished by dust having an
abnormal extinction law, perhaps with $R_{V} \approx 1.5$ (Wang et
al. 2008a). Applying this extinction correction (which is an
extrapolation from the results of Wang et al. 2008a) and the red-tail
correction (which is an extrapolation from the results of Brown et
al. 2010) to the UV magnitudes of SN 2006X would make it appear
brighter than the other comparisons by $\sim 1.7 \pm 0.6$ mag in
uvw1/F250W. We caution, however, that this result may suffer large
uncertainties from the speculated extinction and red-tail corrections.

\section{Discussion}

\subsection{Bolometric Light Curve and Nickel Mass}
\begin{figure}
\figurenum{11}
\vspace{-0.6cm}
\hspace{-0.7cm}
\includegraphics[angle=0,width=100mm]{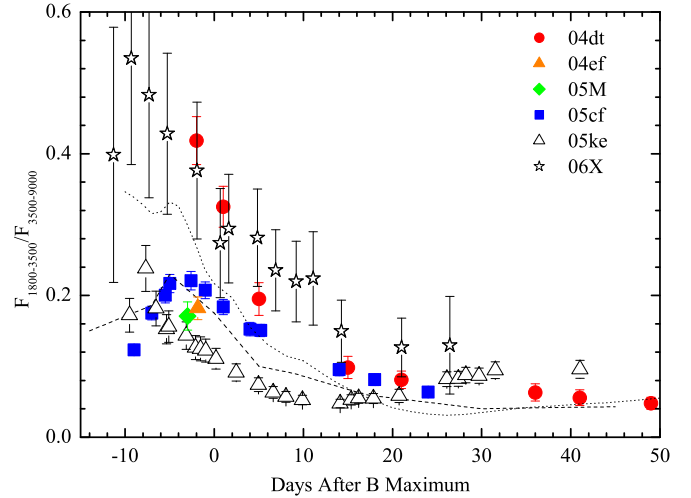}
\vspace{-0.5cm}
\caption{Flux ratio of the UV (1800--3500~\AA) portion to the optical
  flux (3500--9000~\AA), measured from the flux-calibrated UV-optical
  spectra of SNe 2004dt, 2004ef, 2005M, 2005cf, 2005ke, and 2006X (see
  text for the sources of the data). The dotted line is the flux ratio
  computed from the template spectra from Hsiao et al. (2007); the
  dashed curve is the flux ratio estimated from a few nearby SNe~Ia
  with the UV observations (Stanishev et al. 2007).}
\label{fig-11}
\end{figure}

To examine how the relative flux in the UV evolves after the SN
explosion, we compute the ratio of the UV emission (1800--3500~\AA) to
the optical (3500--9000~\AA), $F_{\rm UV}/F_{\rm opt}$, for a few
SNe~Ia, as shown in Figure 11. The flux ratios obtained for SN 2005cf
and SN 2005ke are overplotted. The dotted curve represents the Hsiao
et al. (2007) template and the dashed curve shows a combined template
of SNe 1981B, 1989B, 1990N, 1992A, and 2001el (Stanishev et
al. 2007). The flux ratios were calculated by integrating the
flux-calibrated, UV-optical spectra except for SN 2005ke and SN
2006X. The integrated fluxes of these two SNe~Ia are obtained
approximately by the mean flux multiplied by the effective width of
the passband. It is noteworthy that the flux ratio shown for SN 2006X
might be just a lower limit, as we did not include the flux
contribution below 2500~\AA\ in the analysis because of the larger
uncertainties in the extinction, $K$-corrections, and red-tail
corrections at shorter wavelengths (e.g., Brown et al. 2010).

We notice that the $F_{\rm UV}/F_{\rm opt}$ value of SN 2006X peaks at
$t \approx 10$ days before the $B$-band maximum, about 5 days earlier
than for SN 2005cf and the templates. This feature is similarly
observed in the fast-decliner SN 2005ke, which exhibited a possible
signature of X-ray emission as well as evidence for a UV excess in the
early nebular phase, perhaps suggestive of circumstellar interaction
(Immler et al. 2006). Owing to the lack of earlier UV data, we cannot
conclude whether such a feature is present in SN 2004dt. It is
apparent that the flux ratio $F_{\rm UV}/F_{\rm opt}$ exhibits a large
scatter at comparable phases for the selected sample of
SNe~Ia. Consequently, the spread in bolometric corrections may affect
significantly the inferring of the nickel masses from the light-curve
peaks.

\begin{figure}
\figurenum{12}
\vspace{-0.8cm}
\hspace{-0.7cm}
\includegraphics[angle=0,width=95mm]{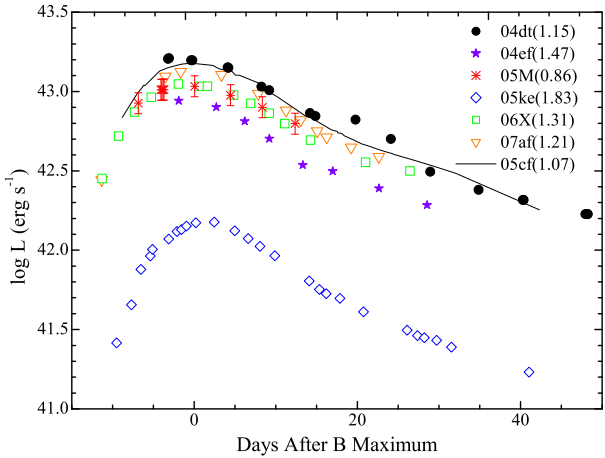}
\vspace{-0.7cm}
\caption{The UV/optical/IR quasi-bolometric light curves of SNe
  2004dt, 2004ef, 2005M, and 2005cf.  Overplotted are the
  corresponding light curves of the comparison SNe~Ia. The numbers in
  parentheses represent the $\Delta m_{15}(B)$ values for the SNe~Ia.}
\label{fig-12}
\end{figure}

The overall properties of the SNe can be represented by their
quasi-bolometric (UV/optical/IR) light curves shown in Figure 12. The
near-infrared (NIR) photometry of SNe 2004ef and 2005M was taken from
Contreras et al. (2010).  The NIR emission of SN 2004dt was corrected
on the basis of SN 2005cf (W09b).  Similar corrections were applied to
the comparison SNe~Ia when the NIR observations were not
available. Overall, the quasi-bolometric light curves of our SNe~Ia
are very similar in shape, with the exception of SN 2004dt, which
shows a prominent ``bump'' feature. This ``bump'' is consistent with the
secondary maximum visible in the $I$ band (see Fig. 2), being more
prominent and occurring 10 days earlier than that of the other
comparisons. This observed behavior suggests that SN 2004dt may have a
larger, cooler iron core, or a higher progenitor metallicity according
to the study of the physical relation between a supernova's NIR
luminosity and its ionization state (Kasen 2006).

The bolometric luminosity at maximum light of SN 2004dt is estimated
to be $L_{\rm max} \approx (1.7 \pm 0.2) \times 10^{43}$ erg s$^{-1}$
with $R_{V} = 2.3$ and $E(B-V)_{\rm host} = 0.05$ mag.  This value is
consistent with that obtained for SNe~Ia with similar $\Delta m_{15}$
such as SN 2005cf [$\sim (1.5 \pm 0.2) \times 10^{43}$ erg s$^{-1}$] and
SN 2007af [$\sim (1.3 \pm 0.2) \times 10^{43}$ erg s$^{-1}$] within
1--2$\sigma$ errors.  With the derived peak luminosity, we can
estimate the $^{56}$Ni mass ejected during the explosion (Arnett
1982). According to Stritzinger \& Leibundgut (2005), the $^{56}$Ni
mass ($M_{\rm Ni}$) can be written as a function of the bolometric
luminosity at maximum and the rise time $t_{r}$:
\begin{equation}
L_{\rm max} = (6.45e^{-t_{r}/(8.8~{\rm d})} +
1.45e^{-t_{r}/(111.3~{\rm d})})(\frac{M_{\rm
Ni}}{{\rm M}_{\odot}})\times 10^{43} \, \rm{erg\ s^{-1}.}
\emph{}\end{equation}
\noindent
Taking the rise time $t_{r}$ as $\sim 20$~d for SN 2004dt and SN
2005cf, $\sim 18$~d for SN 2004ef, and $\sim 24$~d for SN 2005M (e.g.,
Ganeshalingam et al. 2011), the corresponding mass of the ejected
$^{56}$Ni is roughly estimated to be $0.9 \pm 0.1$ M$_{\odot}$, $0.8
\pm 0.1$ M$_{\odot}$, $0.4 \pm 0.1$ M$_{\odot}$, and $0.7 \pm 0.1$
M$_{\odot}$, respectively.

We see that the deduced $M_{\rm Ni}$ of the four SNe~Ia shows
significant scatter, and the value for SN 2004dt is slightly larger
than the normal values. The net effect of a larger $M_{\rm Ni}$ would
tend to delay the secondary maximum of the NIR light curves,
inconsistent with what is seen in SN 2004dt. This inconsistency can be
resolved if SN 2004dt has a shorter rise time, $t_{r} \approx
18.0$~d. Given the observational evidence that longer $t_{r}$ usually
corresponds to slower post-maximum decline of the light curve
(Ganeshalingam et al. 2011), SN 2004dt may rise to maximum at a faster
pace. A shorter rise time is likely to be a common feature of the
high-velocity SNe~Ia (Zhang et al. 2010; Ganeshalingam et
al. 2011). Therefore, the deduced nickel mass of SN 2004dt may have
been overestimated to some extent. On the other hand, the implied
$M_{\rm Ni}$ for SN 2005M seems quite normal and is lower than in
slowly declining SNe~Ia such as SN 1991T, indicating that the
light-curve widths are not a single-parameter family of the ejected
nickel mass.

\subsection{Origin of the UV Excess in SN 2004dt}

The UV emission of SNe~Ia is thought to originate predominantly from
the outer layers of the ejecta because UV photons produced in deeper
layers are subject to line blanketing by the wealth of bound-bound
transitions associated with iron-group elements. Therefore, UV
features are a promising probe to study the composition of the outer
ejecta of SNe~Ia.

\begin{figure*}[!ht]
\figurenum{13}
\vspace{-1.3cm}
\hspace{-0.5cm}
\includegraphics[angle=0,width=190mm]{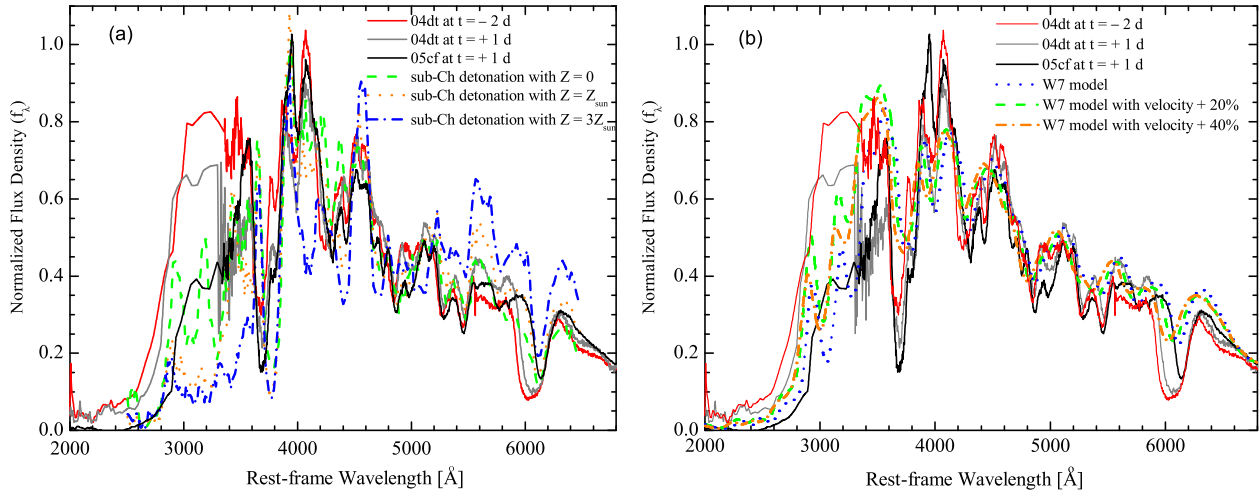}
\vspace{-1.2cm}
\caption{Left panel: Comparison of the UV spectra of SN 2004dt near
  maximum light with synthetic UV spectra produced at a comparable phase
  by the sub-Chandrasekhar detonation model with different
  metallicity. Right panel: Similar comparison with the synthetic
  spectra, but obtained with the W7 model and different expansion
  velocities.}
\label{fig-13}
\end{figure*}

Previous work has shown that the UV is particularly sensitive to the
metal content of the outer ejecta (line-blocking effect) and their
ionization (backwarming effect: Lentz et al. 2000; Sauer et
al. 2008). Lentz et al. (2000) suggested a correlation between the
emitted UV flux and the progenitor metallicity, with lower flux for
models with higher metallicity. On the other hand, Sauer et al. (2008)
proposed that the UV flux can become stronger at high metallicity in
the outer layers due to an enhanced reverse-fluorescence scattering of
photons from red to blue wavelengths and a change in the ionization
fraction (backwarming effect). According to the model series for SN
2001eh and SN 2001ep (Sauer et al. 2008), as well as the prediction by
Lentz et al. (2000), an abundance change of about $\pm 2.0$ dex (e.g.,
variations from 1/10 normal metallicity in the C+O layer to 10 times
normal metallicity in the C+O layer) could lead to a change of up to
$\sim 0.3$ mag in F250W/uvw1. This is much smaller than that observed
in SN 2004dt.  Note that the above studies are based on the
one-dimensional deflagration model W7 (Nomoto et~al. 1984).

To illustrate the effect of different progenitor metallicity on the UV
spectra in more detail, we also consider a sub-Chandrasekhar-mass
detonation model proposed by Sim et al. (2010). In particular, we
adopt the model of a $1.06~{\rm M}_\odot$ CO white dwarf (model 1.06),
which was found to give good agreement with the observed properties of
normal SNe~Ia in the optical. For this model, Sim et al. (2010)
studied the effect of progenitor metallicity on the nucleosynthetic
yields and, in turn, the synthetic observables.  For that purpose they
polluted the initial CO white dwarf with 7.5\% $^{22}$Ne
(corresponding to a progenitor metallicity of $\sim 3\,{\rm
  Z}_{\odot}$). From this they find that the metallicity mainly
affects the blue bands (specifically, they find the $B$-band peak
magnitude to be $\sim0.5$\,mag dimmer for the model with $Z=3{\rm
  Z}_{\odot}$ compared to the model with $Z=0$). In Figure~13(a), we
show the UV spectra of these models at maximum light, which were
obtained using Monte Carlo radiative transfer code ARTIS (Kromer \&
Sim 2009).  To get better coverage in progenitor metallicity, we added
another model at $Z={\rm Z}_{\odot}$ which was obtained in exactly the
same manner as described by Sim et~al. (2010). It is clearly seen that
the UV flux increases significantly with decreasing progenitor
metallicity. However, comparing the synthetic spectra to observed
spectra of SN~2004dt, it is also obvious that the metallicity effect
cannot be the dominant factor responsible for the unusual brightness
of SN 2004dt in the UV, as it is not possible to enhance the UV flux
farther than for the model with zero metallicity.

SN 2004dt shows line-expansion velocities which are apparently larger
than those of normal SNe~Ia. To investigate if these higher ejecta
velocities can explain the UV excess of SN 2004dt, we adopted the
standard deflagration model W7 (Nomoto et al. 1984), which is known to
reproduce the observed maximum-light spectra of SNe~Ia very well. The
$t \approx 0$ day models shown in Figure~13(b) were calculated using
the time-dependent PHOENIX code in local thermodynamic equilibrium
(LTE; Jack et al. 2009; 2011). The velocities in W7 were increased by
a uniform factor (20\% or 40\%). The densities were adjusted so that
the total mass was conserved. Figure~12(b) clearly shows enhanced UV
emission when the velocities are larger; the effects of line blending
in the UV serve to increase the radius of the UV
pseudo-photosphere. Normalizing the spectral flux over the
4000--5500~\AA\ region, we find that the UV flux emitted in the
2500--3500~\AA\ region can be increased by about 40\% if the expansion
velocity $v_{\rm exp}$ is increased by 20\% everywhere in W7. Note
that increasing $v_{\rm exp}$ by 40\% does not further enhance the UV
flux, and the resulting flux increase in the 2500--3500~\AA\ region
drops to 34\%. This indicates that an increase of the expansion
velocity can lead to more UV flux, but it cannot reproduce the very
large flux enhancement near 3000~\AA\ in SN 2004dt.

Interestingly, SN 2004dt is the most highly polarized SN~Ia ever
observed. Across the Si~II line its polarization $P_{\rm Si~II}$
reaches up to 1.6\% at $\sim 1$ week before maximum light (Wang
et~al. 2006), indicating that its Si~II layers substantially depart
from spherical symmetry. Among the comparison sample, SN 2006X also
has a large degree of polarization in the early phase, with $P_{\rm
Si~II} \approx 1.1$\% at 6~d before maximum light (Patat
et~al. 2009). Thus, it could well be that viewing-angle effects play a
major role in the observed UV excess of SN 2004dt. Breaking of
spherical symmetry in the explosion is also thought to be a critical
factor responsible for the observed scatter among SNe~Ia (e.g.,
Wang, Baade, \& Patat 2007; Kasen et~al. 2009; Maeda et al. 2010; Maund
et~al. 2011).

Kromer \& Sim (2009) recently studied the effect of asymmetric ejecta
on the light curves and spectra of SNe~Ia using an ellipsoidal (prolate)
toy model. They found that SNe observed along the equator-on axis are always
brighter than those observed along the pole-on axis. This effect is
strongest in the bluer bands because the photons at short wavelengths
are more strongly trapped than photons in other bands and therefore
tend to preferentially leak out along the equatorial plane where the
photospheric velocity is smallest. Around maximum light, the difference
$\Delta M = M_{\rm pole} - M_{\rm equator}$ measured in the $U$ band
could be $\sim 0.4$ mag larger than that in the $V$ band (Kromer \&
Sim 2009). After maximum light, this line-of-sight effect becomes
weaker with time as the ejecta become optically thin at these
wavelengths. Thus, the UV excess seen in SN 2004dt could also be due
to a geometric effect.

We have described how there are several competing possible
explanations for the enhanced UV flux in SN~2004dt.  This is due to
the complex nature of line blanketing in the barely optically thick,
but highly scattering dominated, differentially expanding SN~Ia
atmosphere.  This effect was described in part by Bongard et
al. (2008), who showed that the complete spectrum is formed throughout
the semitransparent atmosphere and that Fe~III lines produce features
that are imprinted on the full spectrum, partially explaining the UV
excess found in SN~2004dt. Indeed, the differing results of
H\"{o}flich et al. (1998) and Sauer et al. (2008), compared with those
of Lentz et al. (2000), are perhaps due to the complex nature of
spectral formation, ionization, and radiative transfer effects.

While asymmetry could play some role, there seems to be enough
variation in spherically symmetric models due to effects of varying
ionization that is likely produced by a density profile that differs
from that of W7. Recall that W7 has a density bump that occurs when
the flame is quenched and momentum conservation causes material to
pile up ahead of the dying flame. Also, the higher brightness and
indications of a somewhat higher nickel mass in SN~2004dt will affect
the ionization state of the iron-peak elements, which are almost
certainly responsible in part for the observed UV excess. More UV
observations of nearby SNe~Ia are needed to thoroughly understand
these effects, and whether they are due to asymmetries or other
phenomena in the explosion.

\section{Conclusions}

We have presented {\it HST} ACS UV photometry and spectra of
SNe~Ia obtained during Cycle 13 (2004--2005).  These data include 34
ACS prism spectra and 110 photometric observations of four SNe~Ia (SNe
2004dt, 2004ef, 2005M, and 2005cf). The spectral analysis is limited
by the low resolution and low S/N of the data. However, comparison
with the existing low-redshift mean spectra of SNe~Ia clearly
indicates that significant dispersion exists at wavelengths below
$\sim 3500$~\AA. In particular, SN 2004dt is found to show a
prominent, broad emission peak at 3000--3500~\AA\ in the near-maximum
spectrum, rather than having an absorption-like feature at $\sim
3250$~\AA\ similar to that in normal SNe~Ia such as SN 2005cf. Another
interesting feature of SN 2004dt is the rapid decline of its UV
emission after the peak.

Based on a larger sample of SNe~Ia, we studied the properties of their
peak luminosity in the UV region. The luminosity in uvw1/F250W shows
a correlation with the light-curve parameter $\Delta m_{15}(B)$, but
with significantly larger scatter than that found in the optical:
$\sim 0.4$ mag in uvw1/F250W vs. $\sim 0.2$ mag in $B$.  The increased
dispersion in the UV has also been noted in other studies with
independent SN~Ia samples of both photometry (e.g., Guy et~al. 2007;
Brown et~al. 2010) and spectra (Ellis et~al. 2008; Cooke et~al. 2011;
Foley et~al. 2012), which is likely intrinsic and has been interpreted
as compositional differences between events. However, variations of
the abundance based either on the W7 model or the
sub-Chandrasekhar-mass detonation model cannot account for the UV
excess seen in SN 2004dt. The W7 model with increased expansion
velocities has also been investigated, but it can explain only part of
the large UV flux in this object.

In our study, the comparison object SN 2006X may also exhibit strong
emission in the UV. Some common features for SN 2004dt and SN 2006X
are the distinctly high-velocity features beyond the photosphere,
slower $B$-band light-curve evolution in the early nebular phase (Wang
et al. 2008a), and a large degree of ejecta polarization near the
optical maximum (Patat et al. 2009). The above observed features
of these two objects are perhaps related to an asymmetric explosion of
the high-velocity SNe~Ia (Wang et al. 2012, in prep.). Line-of-sight
effects due to the asymmetric explosion can have a more significant
effect on the observed scatter in the UV than in other bands. A new proposal
to identify the progenitors based on the symmetry properties of the explosion
has recently been proposed by Livio \& Pringle (2011). However,
more detailed studies are clearly needed to investigate this fully for
realistic models.

The question of whether a UV excess is a more general property of the
high-velocity subclass merits further study with a larger sample
having earlier observations like PTF~11kly/SN~2011fe (Brown et~al. 2011).
We note, however, that the origin of the high-velocity features and large
polarization observed in SN 2004dt and SN 2006X probably differs, since the
former does not follow the relation between the early-phase velocity gradient
and the nebular-phase emission-line velocity shift while the latter does
(Maeda et al. 2010). SN 2004dt is also found to be an outlier in the
relation between the line polarization of Si~II $\lambda$6355 and the
nebular velocity offset (Maund et al. 2010). Moreover, the reddening
of SN 2004dt by its host galaxy is low, while SN 2006X is heavily
extinguished, perhaps by circumstellar dust (Patat et al. 2007; Wang
et al. 2008a). Thus, the physical origin of the UV excess in these
two SNe~Ia might not be the same. In SN 2006X, dust scattering may
contribute in part to the unusually bright behavior in the UV, which
is favored by the detection of surrounding circumstellar material
(Patat et al. 2007) and/or a light echo around this object (Wang et
al. 2008b).

We further emphasize here the significance of the discovery of a UV
excess in SN 2004dt. It provides not only a new clue to the study of
SN~Ia physics and/or the progenitor environments, but also draws
attention to another possible systematic error that might exist in
current cosmological studies: the relatively higher UV flux would
result in a bluer $U-B$ color for some SNe~Ia, which could lead to an
underestimate of the host-galaxy reddening and hence an overestimate
of the distance. It is of interest to determine the fractional
population of the SN 2004dt-like events or those showing a UV
excess in both the local and distant universe, and to estimate the
impact of such a peculiar subclass of SNe~Ia on current cosmological results.

\acknowledgments We thank Mark Sullivan and Andy Howell for their
suggestions. Financial support for this work has been provided by
the National Science Foundation of China (NSFC grants 11178003,
11073013, and 10173003) and the National Key Basic Research Science
Foundation (NKBRSF TG199075402). A.V.F.'s group at U.C. Berkeley is
grateful for the support of NSF grants AST-0607485 and AST-0908886, the
TABASGO Foundation, and US Department of Energy grants DE-FC02-06ER41453
(SciDAC) and DE-FG02-08ER41563. Substantial financial support for this
work was also provided by NASA through grant GO-10877 from the Space
Telescope Science Institute, which is operated by Associated
Universities for Research in Astronomy, Inc., under NASA contract NAS
5-26555. The work of L.W. is supported by NSF grant
AST-0708873. J.C.W. is supported by NSF grant AST-0707769. K.N. is
supported by WPI Initiative, MEXT, Japan.  M.T., S.B., and E.C. are
supported by grant ASI-INAF I/009/10/0. P.A.M. is supported by NASA
ADP NNX06AH85G. The work of M.H. is supported by ICM grant
P10-064-F and CONICYT grants 150100003 and PFB-06, Chile

\clearpage

\begin{deluxetable}{lllcr}
\hspace{-1.0cm}
\tablecolumns{6} \tablewidth{0pc} \tabletypesize{\scriptsize}
\tablecaption{Journal of {\it HST} ACS PR200L Spectroscopic Observations of Type Ia Supernovae}
\tablehead{\colhead{SN} & \colhead{UT Date} & \colhead{JD\tablenotemark{a}} &
\colhead{Phase\tablenotemark{b}} & \colhead{Exp.(s)}}
\startdata
SN 2004dt& 2004 Aug. 20 &3237.8&$-$2.0&350 \\
SN 2004dt& 2004 Aug. 23 &3240.6& +0.8 &350 \\
SN 2004dt& 2004 Aug. 27 &3245.1& +5.3 &350 \\
SN 2004dt& 2004 Aug. 31 &3249.2& +9.4 &350 \\
SN 2004dt& 2004 Sep. 01 &3250.2& +10.4&470 \\
SN 2004dt& 2004 Sep. 06 &3255.2& +15.4&350 \\
SN 2004dt& 2004 Sep. 07 &3255.8& +16.0&490 \\
SN 2004dt& 2004 Sep. 12 &3260.8& +21.0&360 \\
SN 2004dt& 2004 Sep. 16 &3265.1& +25.3&360 \\
SN 2004dt& 2004 Sep. 21 &3269.9& +30.1&360 \\
SN 2004dt& 2004 Sep. 27 &3275.8& +36.0&360 \\
SN 2004dt& 2004 Oct. 02 &3281.3& +41.5&720 \\
SN 2004dt& 2004 Oct. 10 &3289.1& +49.3&1080 \\
         &              &      &      &     \\
SN 2004ef& 2004 Sept. 14 &3263.1&$-$1.5&1050 \\
SN 2004ef& 2004 Sept. 18 &3267.3& +2.7 &1050 \\
SN 2004ef& 2004 Sept. 22 &3270.8& +6.2 &1050 \\
SN 2004ef& 2004 Sept. 25 &3273.9& +9.3 &1050 \\
SN 2004ef& 2004 Sept. 29 &3277.9& +13.3&360  \\
SN 2004ef& 2004 Oct. 02 &3281.4& +17.0&360  \\
SN 2004ef& 2004 Oct. 08 &3287.2& +22.6&240  \\
SN 2004ef& 2004 Oct. 14 &3293.1& +28.5&240  \\
         &              &      &      &     \\
SN 2005M & 2005 Jan. 31 &3402.0&$-$2.8&2760 \\
         &              &      &      &     \\
SN 2005cf& 2005 Jun. 03 &3525.0&$-$8.8&480  \\
SN 2005cf& 2005 Jun. 05 &3527.2&$-$6.6&480  \\
SN 2005cf& 2005 Jun. 07 &3529.3&$-$4.5&480  \\
SN 2005cf& 2005 Jun. 11 &3533.0&$-$0.8&720  \\
SN 2005cf& 2005 Jun. 14 &3536.0& +2.2 &720  \\
SN 2005cf& 2005 Jun. 16 &3537.9& +4.1 &720  \\
SN 2005cf& 2005 Jun. 21 &3542.6& +8.8 &1680 \\
SN 2005cf& 2005 Jun. 25 &3547.2& +13.4&840  \\
SN 2005cf& 2005 Jun. 26 &3547.9& +14.1&840  \\
SN 2005cf& 2005 Jun. 29 &3551.2& +17.4&840  \\
SN 2005cf& 2005 Jun. 30 &3551.7& +17.9&840  \\
SN 2005cf& 2005 Jul. 05 &3557.0& +23.2&1680 \\
\enddata
\tablenotetext{a}{Julian Date minus 2,450,000.}
\tablenotetext{b}{Days relative to the epoch of $B$-band maximum.}
\end{deluxetable}

\begin{deluxetable}{lccrrrrrrrr}
\hspace{-1.0cm}
\tablecolumns{11} \tablewidth{0pc} \tabletypesize{\scriptsize}
\tablecaption{{\it HST} ACS Ultraviolet Photometry of Type Ia Supernovae}
\tablehead{\colhead{UT Date} & \colhead{JD\tablenotemark{a}} &
\colhead{Phase\tablenotemark{b}} & \colhead{F220W} & \colhead{F250W}
& \colhead{F330W} &\colhead{$RC_{\rm F220W}$} &\colhead{$RC_{\rm F250W}$} & \colhead{$K_{\rm F220W}$}
& \colhead{$K_{\rm F250W}$} & \colhead{$K_{\rm F330W}$} \\
\colhead{}&\colhead{} &\colhead{} & \colhead{(mag)} & \colhead{(mag)} & \colhead{(mag)} & \colhead{(mag)}
& \colhead{(mag)} & \colhead{(mag)} & \colhead{(mag)} & \colhead{(mag)}}  \\
\startdata
 &      &          &   &   &  SN 2004dt     &          &    &  &   &       \\
\tableline
2004 Aug. 20 &3237.76   &$-$2.04  & 17.40(03) & 15.72(03) & 14.45(02) &0.03(01) &0.01(01) &0.04(01) &0.11(01) & 0.02(01) \\
2004 Aug. 23 &3240.68   & +0.88   & 17.68(02) & 15.97(03) & 14.65(02) &0.03(01) &0.01(01) &0.05(01) &0.11(01) & 0.04(01) \\
2004 Aug. 27 &3245.15   & +5.35   & 18.16(03) & 16.61(03) & 15.30(02) &0.02(01) &0.01(01) &0.05(01) &0.10(01) & 0.06(02) \\
2004 Aug. 31 &3249.25   & +9.45   & 18.65(03) & 17.24(02) & 15.95(02) &0.02(01) &0.01(01) &0.07(01) &0.10(01) & 0.06(02) \\
2004 Sep. 01 &3250.18  & +10.38  & 18.71(03) & 17.27(02) & 16.09(02) &0.02(01) &0.01(01) &0.07(01) &0.10(01) & 0.06(02) \\
2004 Sep. 06 &3255.18  & +15.38  & 19.25(03) & 18.06(03) & 16.88(02) &0.01(01) &0.01(01) &0.06(01) &0.09(01) & 0.07(02) \\
2004 Sep. 07 &3255.85  & +16.05  & 19.31(03) & 18.22(03) & 17.14(03) &0.01(01) &0.01(01) &0.06(01) &0.08(01) & 0.07(02) \\
2004 Sep. 12 &3260.78  & +20.98  & 19.88(03) & 18.89(02) & 17.89(02) &0.01(01) &0.01(01) &0.06(01) &0.08(01) & 0.07(02) \\
2004 Sep. 16 &3265.11  & +25.31  & 20.08(03) & 19.25(03) & 18.38(02) &0.01(01) &0.01(01) &0.03(01) &0.07(02) & 0.07(02) \\
2004 Sep. 21 &3269.92  & +30.12  & 20.26(03) & 19.59(03) & 18.70(02) &0.01(01) &0.01(01) &0.03(01) &0.06(02) & 0.06(01) \\
2004 Sep. 27 &3275.85  & +36.05  & 20.55(03) & 20.08(03) & 19.07(02) &0.01(01) &0.01(01) &0.04(01) &0.07(02) & 0.06(01) \\
2004 Oct. 02  &3281.34  & +41.54  & 20.88(03) & 20.39(03) & 19.33(03) &0.01(01) &0.01(01) &0.05(04) &0.06(02) & 0.07(01) \\
2004 Oct. 10  &3289.10  & +49.30  & 21.13(03) & 20.55(03) & 19.48(03) &0.01(01) &0.01(01) &0.02(02) &0.06(01) & 0.06(01) \\
\tableline
&       &     &    &       &   SN 2004ef     &       &  &  &    &                        \\
\tableline
2004 Sep. 13 &3262.23   &$-$1.87  & \nodata   & 19.06(03) & 16.98(02) &0.08(07)  &0.02(01) & \nodata  & 0.19(04)&0.07(03) \\
2004 Sep. 18 &3266.83   &+2.94    & 21.62(06) & 19.57(03) & 17.71(02) &0.04(02)  &0.02(01) & 0.22(14) & 0.17(02)&0.10(01) \\
2004 Sep. 21 &3270.32   &+6.41    & 21.85(06) & 19.87(03) & 18.12(03) &0.04(02)  &0.02(01) & 0.18(09) & 0.17(03)&0.13(01) \\
2004 Sep. 24 &3273.33   &+9.41    & 22.15(09) & 20.31(03) & 18.59(03) &0.05(03)  &0.02(01) & 0.15(05) & 0.17(03)&0.13(01) \\
2004 Sep. 28 &3277.43   &+13.54   & 22.12(03) & 20.52(02) & 19.04(02) &0.06(04)  &0.02(01) & 0.08(02) & 0.17(02)&0.13(01) \\
2004 Oct. 02 &3281.08    &+17.13   & 22.58(04) & 20.93(02) & 19.60(02) &0.04(03)  &0.02(01) & 0.19(05) & 0.20(07)&0.12(02) \\
2004 Oct. 08 &3286.77    &+22.88   & \nodata   & 21.27(03) & 20.23(02) &0.02(05)  &0.01(01) & \nodata  & 0.10(04)&0.10(02) \\
2004 Oct. 14 &3292.66    &+28.79   & \nodata   & 21.74(03) & 20.66(03) &0.01(05)  &0.01(01) & \nodata  & 0.09(04)&0.13(02) \\
\tableline
&      &      &   &        &  SN 2005M   &    &     &        &       &             \\
\tableline
2005 Jan. 28 &3398.48   &$-$6.29  & 20.07(02) & 17.54(02)   & 16.02(03) &0.02(02)  & 0.01(01)   &0.13(11) &0.13(03)   &0.01(01) \\
2005 Jan. 31 &3401.60   &$-$3.30  & 19.91(02) & 17.45(04)   & 15.88(02) &0.03(02)  & 0.01(01)   &0.15(10) &0.12(02)   &0.02(01) \\
2005 Feb. 04 &3405.97   &+1.23    & 19.96(02) & \nodata     & 15.93(03) &0.02(02)  & \nodata    &0.06(05) &\nodata    &0.05(01) \\
2005 Feb. 09 &3410.70   &+5.96    & 20.57(03) & \nodata     & 16.51(03) &0.02(02)  & \nodata    &0.06(01) &\nodata    &0.07(02) \\
2005 Feb. 13 &3414.90   &+10.16   & 20.87(03) & \nodata     & 16.85(03) &0.02(02)  & \nodata    &0.08(02) &\nodata    &0.09(01) \\
2005 Feb. 17 &3419.29   &+14.49   & 20.82(03) & \nodata     & 17.35(03) &0.02(02)  & \nodata    &0.07(01) &\nodata    &0.11(03)\\
\enddata
\tablenotetext{a}{Julian Date minus 2,450,000.}
\tablenotetext{b}{Relative to the epoch of $B$-band maximum.}
\tablenotetext{}{Note: uncertainties, in units of 0.01 mag, are $1\sigma$.
See W09b for {\it HST} UV photometry of SN 2005cf.
}
\end{deluxetable}


\begin{deluxetable}{lllcrrrrr}
\hspace{-1.0cm}
\tablecolumns{9} \tablewidth{0pc} \tabletypesize{\scriptsize}
\tablecaption{Relevant Parameters for the {\it HST} UV Sample of Type Ia Supernovae}
\tablehead{\colhead{SN} & \colhead{Host Galaxy} & \colhead{$v_{3~{\rm K},v220}$(km s$^{-1}$)} &
\colhead{$\mu$ (mag)} & \colhead{$\Delta m_{15}(B)$} & \colhead{$B_{\rm max} - V_{\rm max}$} & \colhead{$E(B-V)_{\rm Gal}$} & \colhead{$E(B-V)_{\rm host}$} & \colhead{Source}}
\startdata
SN 2004dt&NGC 0799 &5644&34.41(19)&1.13(04)&0.03(03)&0.020&0.08(04)&1,2,3  \\
SN 2004ef&UGC 12158&8931&35.41(16)&1.46(06)&0.15(03)&0.056&0.11(05)&1,4  \\
SN 2005M &NGC 2930 &6891&34.81(17)&0.86(05)&$-$0.05(04)&0.022&0.10(05)&1,3,5 \\
SN 2005cf&MCG-01-39-003&2143&32.31(34)&1.07(03)&0.09(03)&0.097&0.10(03)&6,7 \\
\enddata
\tablenotetext{}{Note: uncertainty estimates in parentheses are in units of 0.01 mag.}
\tablenotetext{}{1. Ganeshalingam et al. (2010); 2. Altavilla et al. (2007);
3. Contreras et al. (2010); 4. Silverman et al. (2012);
5. Hicken et al. (2009); 6. Wang et al. (2009b); 7. Garavini et al. (2007).}
\end{deluxetable}

\begin{deluxetable}{lllcc}
\hspace{-1.0cm}
\tablecolumns{5} \tablewidth{0pc} \tabletypesize{\scriptsize}
\tablecaption{Linear Fit to the $M_{\rm max} - \Delta m_{15}(B)$ Relation \\
$M_{\rm max} = M_{0} + \alpha~ (\Delta m_{15}(B) - 1.1)$}
\tablehead{\colhead{Bandpass} & \colhead{$M_{0}$ (mag)} &
\colhead{$\alpha$} & \colhead{Number} & \colhead{$\sigma$ (mag)}}
\startdata
\tableline
&   &   $R_{{\rm MW},V} = 3.1$  &  & \\
\tableline
uvw1 & $-$18.18(10) & $-$1.58(39) & 15 & 0.44\\
$U$ & $-$19.88(07) & $-$1.47(27) & 14 & 0.35\\
$B$ & $-$19.42(07) & $-$1.27(24) & 14 & 0.23\\
\tableline
&   &   $R_{V} = 2.3$  &  & \\
\tableline
uvw1& $-$18.12(10) & $-$1.50(39) & 15 & 0.41\\
$U$ & $-$19.67(07) & $-$1.18(27) & 14 & 0.29\\
$B$ & $-$19.23(07) & $-$1.00(24) & 15 & 0.20\\
\tableline
&   &   $R_{{\rm CSLMC},V} = 1.8$  &  & \\
\tableline
uvw1 & $-$18.05(10) & $-$1.38(39) & 14 & 0.40\\
$U$ & $-$19.60(07) & $-$1.08(27) & 14 & 0.28\\
$B$ & $-$19.24(07) & $-$1.01(24) & 14 & 0.20\\
\tableline
\enddata
\tablenotetext{}{Note: uncertainty estimates in parentheses are in
units of 0.01 mag.}
\end{deluxetable}

\end{document}